\RequirePackage{tikz}
\documentclass[a4paper]{article}
\usepackage{hyperref}
\usepackage[utf8]{inputenc}
\usepackage{fullpage}
\usepackage{amsmath,amssymb,mathrsfs}

\newcommand{\Perp}{\perp \! \! \! \perp}
\usepackage{color}

\usepackage[caption=false]{subfig}

\begin{document}

\title{Causal discovery for observational sciences using supervised machine learning}

\author{Anne Helby Petersen\thanks{Corresponding author: ahpe@sund.ku.dk} \\University of Copenhagen \and Joseph Ramsey \\Carnegie Mellon University \and Claus Thorn Ekstrøm \\University of Copenhagen \and Peter Spirtes \\Carnegie Mellon University}

\date{}

\maketitle

\abstract{Causal inference can estimate causal effects, but unless data are collected experimentally, statistical analyses must rely on pre-specified causal models. Causal discovery algorithms are empirical methods for constructing such causal models from data. 

Several asymptotically correct methods already exist, but they generally struggle on smaller samples. Moreover, most methods focus on very sparse causal models, which may not always be a realistic representation of real-life data generating mechanisms. Finally, while causal relationships suggested by the methods often hold true, their claims about causal non-relatedness have high error rates. This non-conservative error trade off is not ideal for observational sciences, where the resulting model is directly used to inform causal inference: A causal model with many missing causal relations entails too strong assumptions and may lead to biased effect estimates. 

We propose a new causal discovery method that addresses these three shortcomings: Supervised learning discovery (SLdisco). SLdisco uses supervised machine learning to obtain a mapping from observational data to equivalence classes of causal models. 

We evaluate SLdisco in a large simulation study based on Gaussian data and we consider several choices of model size and sample size. We find that SLdisco is more conservative, only moderately less informative and less sensitive towards sample size than existing procedures. 

We furthermore provide a real epidemiological data application. We use random subsampling to investigate real data performance on small samples and again find that SLdisco is less sensitive towards sample size and hence seems to better utilize the information available in small datasets.
}

\paragraph{Acknowledgments:} We thank Dan Saattrup Nielsen for guidance and help with the machine learning procedures. We thank Kun Zhang for his contributions to designing the project. This work was funded by the Independent Research Fund Denmark (grant 8020-00031B).

\section{Introduction}\label{sec1}

Questions of cause and effect are prevalent in many scientific applications: Why is the global temperature rising? What are the economic consequences of raising a tax? How can we prevent the development of depression? Classic scientific methodology suggests that such questions ought to be answered by relying on experimentation, controlling the supposed cause in order to quantify its effect. However, many interesting and important questions do not allow for such experimentation -- it may be impossible, too costly or not ethically permissible. In such cases, empirical sciences generally rely on theory for constructing causal models and use \textit{observational data} for causal inference, i.e. causal effect estimation. While this methodology has provided many useful results, relying on theoretical specifications of causal models induces risk of confirmation bias and greatly limits the scope of topics that can be studied. 

\textit{Causal discovery} provides an empirical alternative to theory-driven causal model specification. Here, causal models are inferred from observational data by use of conditional independence tests or model scoring. Exhaustive searches among candidate causal models is generally not computationally feasible, and hence, existing causal discovery algorithms rely on either sequential testing, greedy search procedures or a combination of both. While several existing algorithms have been proven to be asymptotically correct (including the Peter-Clark (PC) algorithm \cite{kalisch2007} and greedy equivalence search (GES) \cite{chickering2002}), their finite sample properties are compromised by the sequential nature of the procedures. Each algorithmic step involves one or more statistical decisions, which will then inform subsequent steps. But statistical decisions are of course subject to statistical error, and these statistical errors propagate in an unfortunate manner, which makes the algorithms questionable for small or moderate sample sizes. 

Moreover, existing algorithms have generally been developed and evaluated with focus on sparse causal models, that is, causal models where there are only few cause-effect links between the involved variables. This makes computations simpler, but sparsity of causal mechanisms may not be a meaningful or natural assumption in many sciences relying on observational data for causal inference. For example in epidemiology, causal etiological models often include numerous complexly intertwined cause-effect relationships that can hardly be described as sparse \cite{destavola2006}. 

Finally, causal discovery methods have traditionally been evaluated with focus on their performance for \textit{positive} findings: It has been prioritized to ensure that causal mechanisms claimed to be present are truly present. This focus has been useful for informing experimental sciences where such claimed causal relationships can then afterwards be tested and quantified using controlled experiments. However, in truly observational sciences, such as epidemiology, where no interventions are feasible, this evaluation metric may not be as useful. A major concern in epidemiology is the presence of \textit{confounding} and \textit{selection} variables for a given causal inference task and therefore, a causal discovery procedure that provides good guarantees on \textit{negative} findings may be more useful: Knowing that a certain causal effect is for example \textit{not} confounded will help inform what statistical procedures can be used to estimate it, and avoids conditioning on variables that will bias the estimation of this target (i.e. inadvertently conditioning on mediators or colliders). 
 
To address these problems, we propose a new approach for causal discovery using supervised machine learning, supervised learning discovery (SLdisco). We train a machine learning model on simulated data, where the true causal model is known, thereby obtaining a classification function that takes in new observational data and outputs a causal model suggestion. SLdisco is non-sequential in the sense that it learns the full causal model jointly, and hence it is less sensitive to statistical error occurring from small or moderate sample sizes. It seeks to estimate the true model density, and hence it has no built-in preference towards sparse (or dense) causal model. This agnosticism gives it relatively better performance on dense data generating mechanisms. Finally, the construction of the classification function includes a classification threshold, which will provide a direct means to control trade off between different error types, and hence SLdisco can be optimized to focus on correctly identifying negative rather than positive causal relationships. 

Our proposed method assumes no latent confounding or selection variables, and seeks to learn the Markov equivalence class of the causal model, as represented via a completed partially directed acyclic graph (CPDAG). This article is the first to propose to use supervised machine learning for this task. We use neural networks as the machine learning model and we train these models on simulated linear Gaussian data. The choice of linear Gaussian data is theoretically attractive, since several existing classical causal discovery algorithms are asymptotically correct for this type of data.  

The article is structured as follows: We first introduce key concepts in Section \ref{sec.terminology}. In Section \ref{sec.causaldiscofromobs} we provide a thorough exposition of the three issues mentioned above for two popular existing causal discovery methods, the PC and greedy equivalence search (GES) algorithms. In Section \ref{sec.sldisco} we present our proposed methodology, SLdisco. 
We evaluate SLdisco in a large simulation study in Section \ref{sec.simstudy}, and in a real epidemiological data application in Section \ref{sec.appli}. Finally, we discuss strengths and limitations of our method in Section \ref{sec.discussion} and provide suggestions for future research. 

\subsection{Related work}

This article is the first to propose using supervised machine learning for CPDAG discovery. However, other authors have proposed to use supervised machine learning for other causal discovery tasks.

\cite{li2020} looks at the special case where the full DAG is identifiable from observational data (so-called linear non-Gaussian additive noise (LiNGAM) setting and the linear Gaussian equivariance setting), and applies a neural network to this task. However, they do not consider influence of sample size or graph density, and their evaluation setup does not inform non-conservative performance. They specifically consider only very sparse graphs and all graphs in each of their training datasets have the same density. Hence their results are conditional on knowing the correct graph density. This is not a realistic assumption in observational sciences.  

\cite{yu2019} and \cite{xu2021} also propose methods using neural networks in the same special case where the full DAGs is identifiable, and they also consider only a fixed average graph density, and do not study the influence of sample size. 

\cite{ke2022} also address this special case using neural networks, and they furthermore train their model on a mixture of observational and experimental data. Hence, in observational sciences, where experimental data is not available, this method cannot be used. 

\cite{zheng2018} and \cite{zheng2020} tackle the same special case of identified DAGs, but provide a general framework for formulating the graphical acyclicity criterion as a constraint for a continuous optimization problem. 

Finally, \cite{goudet2018} proposes a neural network based method for learning the directions of causal relationships. Their method assumes that the graph skeleton is known and then seeks to orient each edge by looking only at pairs of adjacent variables. While this avoids error propagation, it does not make use of the full information in the data, and the graph skeleton would still have to be learned with other approaches. 

In conclusion, while the idea of using supervised machine learning in causal discovery is not new, no other work use this methodology to infer CPDAGs; only DAG discovery and orientation classification has previously been addressed. This also means that the models described above cannot be used for CPDAG discovery and hence are not suitable benchmarks for our task.

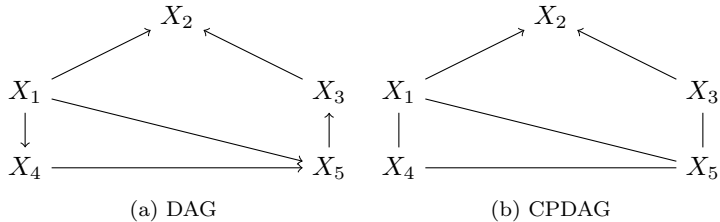
\begin{figure}
\centering
\subfloat[DAG \label{subfig.dag}]{
\begin{tikzpicture}
\node (1) at (0,1) {$X_1$};
\node (2) at (2,2) {$X_2$};
\node (3) at (4,1) {$X_3$};
\node (4) at (0,0) {$X_4$};
\node (5) at (4,0) {$X_5$};
\draw [->] (1) edge (2);
\draw [->] (1) edge (5);
\draw [->] (1) edge (4);
\draw [->] (3) edge (2);
\draw [->] (4) edge (5);
\draw [->] (5) edge (3);
\end{tikzpicture}
}
\subfloat[CPDAG\label{subfig.cpdag}]{
\begin{tikzpicture}
\node (1) at (0,1) {$X_1$};
\node (2) at (2,2) {$X_2$};
\node (3) at (4,1) {$X_3$};
\node (4) at (0,0) {$X_4$};
\node (5) at (4,0) {$X_5$};
\draw [->] (1) edge (2);
\draw [-] (1) edge (5);
\draw [-] (1) edge (4);
\draw [->] (3) edge (2);
\draw [-] (4) edge (5);
\draw [-] (5) edge (3);
\end{tikzpicture}
} 
\caption{An example of a DAG (a) and the corresponding CPDAG (b) which describes its Markov equivalence class.}
\label{fig.dag.cpdag}
\end{figure}

\section{Representing causal models} 
\label{sec.terminology}
\paragraph{DAGs and CPDAGs} 
We focus on causal data generating mechanisms that can be described by directed acyclic graphs (DAGs). We will here only introduce the concepts informally and refer to Appendix \ref{appendix.terminology} for formal definitions and an overview of notation. We will use $\mathbf{X} = X_1, ..., X_p$ to refer to a set of $p$ random variables. In a causal DAG, nodes represent random variables, and edges represent causal links. If $X_i \to X_j$ in a DAG, $X_i$ is a cause of $X_j$, or, equivalently, intervening on $X_i$ will result in changes in $X_j$ (but not vice versa). Figure \ref{subfig.dag} provides an example of a DAG. Causal models that can be represented by DAGs have two important assumptions: First, we assume acyclicity, which means that there cannot be any feedback loops in the data generating mechanism. 
Secondly, we assume that there are no unmeasured variables that are parents of more than one variable (no latent confounders), nor any unobserved selection (no conditioning on latent colliders).

The causal statements of a DAG can be related to conditional independence via the Markov property, which allows us to use the graphical relation $d$-separation to read off conditional independencies: If a set $S \in \mathbf{X} \setminus \{X_i, X_j\}$ $d$-separates $X_i$ and $X_j$, then it holds that $X_i \Perp X_j \mid S$. In order to draw causal inference from observational data, we generally need the converse implication as well, namely that conditional independence implies causal unrelatedness. We refer to this as the \textit{faithfulness assumption}, and equipped with both the Markov property and the faithfulness assumption, we obtain a crucial link from testable statistical properties (conditional independence) to causal statements. However, the full DAG is generally not uniquely identified from conditional independencies, as more than one DAG can lead to the same conditional independence statements. All DAGs that entail the same conditional independence statements comprise an equivalence class, usually denoted the \textit{Markov equivalence class}, and can be represented by a completed partially directed acyclic graph (CPDAG) \cite{peters2017}. A CPDAG has directed edges wherever all members of the equivalence class agree on the orientation of an edge, and undirected edges whenever there is ambivalence about the orientation of the edge. All edge orientations in a CPDAG are related to its \textit{v-structures} \cite{pearl2009}, that is, triplets of nodes $X_i \leftarrow X_j \rightarrow X_k$ where $X_i$ and $X_k$ are not adjacent. Note that all DAGs in a Markov equivalence class share the same adjacencies, also known as the graph skeleton \cite{pearl2009}. Hence a CPDAG will have the same skeleton as any of its member DAGs. Figure \ref{fig.dag.cpdag} provides an example of a DAG and the CPDAG that represents its Markov equivalence class. 

\paragraph{Adjacency matrices} 
DAGs and CPDAGs can also be represented by \textit{adjacency matrices} instead of their graphical objects. A DAG with $p$ nodes is represented by a $p \times p$ matrix $M$ where the $(i,j)$th element is 1 if and only if there is an edge $X_j \to X_i$ in the DAG. For CPDAGs, which also have undirected edges, the adjacency matrix is defined as follows:
\begin{align*}
M[i,j] = 0 \text{ and } M[j,i] = 1 &\quad \Leftrightarrow \quad  X_i \rightarrow X_j   \\
M[i,j] = 1 \text{ and } M[j,i] = 0 &\quad \Leftrightarrow \quad  X_i \leftarrow X_j  \\
M[i,j] = 1 \text{ and } M[j,i] = 1 &\quad \Leftrightarrow \quad  X_i \relbar X_j   \\
M[i,j] = 0 \text{ and } M[j,i] = 0 &\quad \Leftrightarrow \quad  X_i \,\quad\, X_j   \\
\end{align*}
As examples, the adjacency matrices for DAG and CPDAG in Figure \ref{fig.dag.cpdag} are given by $M_1$ and $M_2$ below, respectively:
\begin{equation*}
M_1 = \begin{bmatrix}
0 & 0 & 0 & 0 & 0 \\
1 & 0 & 1 & 0 & 0 \\
0 & 0 & 0 & 0 & 1 \\
1 & 0 & 0 & 0 & 0 \\
1 & 0 & 0 & 1 & 0 
\end{bmatrix} \quad \text{ and } \quad
M_2 = \begin{bmatrix}
0 & 0 & 0 & 1 & 1 \\
1 & 0 & 1 & 0 & 0 \\
0 & 0 & 0 & 0 & 1 \\
1 & 0 & 0 & 0 & 1 \\
1 & 0 & 1 & 1 & 0 
\end{bmatrix}
\end{equation*}

\section{Observational science challenges for selected existing causal discovery procedures}
\label{sec.causaldiscofromobs}

A \textit{causal discovery procedure} takes in a dataset and outputs a representation of its causal data generating mechanism. Unless additional assumptions about the data generating mechanism are made (e.g., distributional assumptions or assumptions about its functional form), only the CPDAG is identifiable. We will now present the three observational science causal discovery challenges  in the context of two specific algorithms, PC and GES. These algorithms are chosen because they represent the two main paradigms within causal discovery: Constraint-based methods (PC) and score-based methods (GES). Constraint-based methods use conditional independence tests to recover the true CPDAG, while score-based methods search through candidate CPDAGs and score them. Both methods rely on numerous sequential statistical decisions in a manner that is problematic for small sample performance and induces a bias towards sparse models, as we will explain further below. 

\subsection{Small sample and dense graph challenges for the PC algorithm}
The PC algorithm \cite{spirtes1991} reconstructs a causal CPDAG from observational data as follows: First, it starts with a completely connected graph over the nodes. It then seeks to prune away edges by looking for \textit{separating sets}. For each adjacent pair of variables, $X_i$ and $X_j$, the algorithm searches for a set $S$ among nodes that are adjacent to $X_i$ or $X_j$ such that $X_i \Perp X_j \mid S$. If such a (possibly empty) set exists, the edge between $X_i$ and $X_j$ is removed. The algorithm will first consider small separating sets (starting with $S = \emptyset$, i.e., marginal independence), and then, if necessary, increase the size of $S$. Starting with small candidates for $S$ is attractive in the interest of efficient and precise computations, as conditional independence tests for larger $S$ are both more computationally expensive and require more statistical power. When all possible edge pruning steps have been carried out, the algorithm orients edges by applying a complete list of orientation rules, which make use of 1) $v$-structure properties, and 2) the assumption of acyclicity. 

In empirical applications, the decisions about conditional independence rely on statistical tests. No generally valid, uniformly consistent test for conditional independence exists \cite{shah2020}, but under certain distributional assumptions, valid tests are available. For example, for jointly Gaussian variables, conditional independence can be asserted by testing for vanishing partial correlations. Given a valid conditional independence test, it has been proven that PC is asymptotically correct \cite{spirtes1991}. 

We wish to highlight two important characteristics of this algorithm that explain why PC struggles with the first two of the three issues mentioned above, namely small sample performance and dense graph performance:
\begin{enumerate}
\item The algorithm conducts a large number of sequential tests, and the results of former tests inform which tests will later be conducted. If for example a test reveals that two variables, $X_i$ and $X_j$ are separable, the edge between them is removed, and they are then no longer adjacent and will not be considered again for future separation attempts. Moreover, because only variables connected with $X_i$ or $X_j$ are considered for candidate separating sets, this also has consequences for future tests involving the two variables: If $X_i$ and $X_j$ are considered separable, and hence the edge between them is removed, $X_j$ will never again be considered to be included in separating sets for $X_i$. But if the original test conclusion was faulty, for example due to statistical error, such an error may then propagate through its influence on subsequent tests. A similar problem occurs in the orientation step.  
\item The algorithm is biased towards sparse graphs. In dense graphs, separating sets will often be large, as there are many possible paths between two given variables that must all be closed. As the size of the separating set increases, the relative statistical power decreases, and hence the type II error increases, which means that the probability of falsely concluding conditional independence between any two variables is larger. This will lead to removing too many edges and hence PC will be biased towards sparse graphs. 
\end{enumerate}
These two characteristics explain why PC may not perform ideally on small or moderate sample sizes (due to error propagation), nor on dense graphs (due to preference towards sparse solutions).

\subsection{Small sample and dense graph challenges for the GES algorithm}
The greedy equivalence search (GES) \cite{chickering2002} algorithm reconstructs a causal CPDAG from observational data as follows: First, it starts with an empty graph, where no nodes are connected. Then, a forward selection type step is carried out: Edges are greedily added until no further edge additions result in increasing a chosen score. Afterwards, a backwards elimination type step is conducted: Edges are greedily removed until no further edge removals result in increasing the score. Finally, the procedure seeks to further increase the score by considering single-edge orientation reversals. 

In terms of the statistical error propagation and sparse graph preference issues, we make the following remarks regarding GES:
\begin{enumerate}
\item The greedy search strategy implies that former statistical decisions (score comparisons) affect subsequent statistical decisions. Hence statistical errors propagate. 
\item Using an empty graph as starting point for a greedy search results in a preference towards sparsity on small samples. On small samples, statistical error will be larger, and this may result in a scoring function with several local optima, even if the scoring function is convex as the sample size tends to infinity.
If such a local optimum exists, the greedy search procedure of GES may get stuck in this optimum. Because sparse graphs are considered first, there is an increased risk of such local optimum being systematically too sparse. 
\end{enumerate}
In conclusion, as for PC, on small or moderate samples, GES will tend to propagate errors and have a built-in preference towards sparse graphs.

\subsection{Error trade off: A need for conservative discovery}
\label{sec.errortradeoff}

Any causal discovery procedure based on empirical data will of course be subject to some amount of statistical error. We will now discuss what type of error is most critical for observational sciences, using epidemiology as a guiding example, and contrast that to the error trade offs inherent in PC and GES. 

In epidemiology, the typical causal inference pipeline involves two steps before an actual data analysis is initiated: \cite{greenland1999}
\begin{enumerate}
\item Construct causal model (often by drawing a DAG)
\item Analyze DAG to assert whether the causal estimand of interest (e.g., an average treatment effect) is identifiable from observational data and with which analytical approach and which epidemiological study design.
\end{enumerate}
Importantly, what analytical approach is appropriate is dictated by the structure of the DAG. Which effects are readily identifiable from a given DAG or CPDAG can be determined using for example Pearl's \textit{do-calculus} \cite{pearl2009} (for DAGs) or Perković's \textit{causal identification formula} \cite{perkovic2020} (for CPDAGs). And if the causal effect of interest is identifiable, an appropriate causal inference method can then be applied to obtain an effect estimate, for example regression adjustment, inverse probability weighting, the G-formula or targeted minimum loss estimation. But even if a causal effect of interest first appears unidentified, the epidemiological toolbox includes another remedy: Study designs. By making use of auxiliary variables, natural experiments or stratification among the observations, it may be possible to make otherwise unidentifiable effects estimable. Examples of such designs are instrumental variable designs, sibling comparison designs, difference-in-difference designs and case-control designs. 

When asserting whether a certain analytical approach or design is appropriate given a DAG (or CPDAG), \textit{absent} adjacencies are crucial. For example, an absent adjacency may make it appear as if a certain effect of interest is unconfounded, which will make it appear identifiable. 
On the other hand, a spurious adjacency in the CPDAG may make actually identifiable effects appear unidentified. But as long as the spurious adjacency is undirected, it will not induce biased effect estimates.

Hence, if a theory-based causal model is replaced by a causal discovery estimate as the first step, we would prefer a \textit{conservative} estimate of the causal model: The model should rather have too many than too few adjacencies, having the estimated graph skeleton be a supergraph of the true skeleton,  and it should rather have too many undirected edges than too many falsely oriented edges. Using a conservative CPDAG estimate will also imply a conservative causal identification analysis: We may not successfully identify effects that \textit{are} present, but unidentified effects will \textit{not} appear identifiable. For example, if an effect is unconfounded in the estimated causal model, it will generally also be unconfounded in the true causal model. But if it is confounded in the estimated causal model, it may or may not be confounded in the true model.

Since PC and GES are asymptotically correct and have a preference towards sparse graphs on finite samples, they will generally \textit{not} estimate a supergraph of the true skeleton and hence they will not be conservative. On the contrary, they were specifically designed to aid pre-experimental screening where the causal discovery algorithm is used to identify the strongest causal relationships in a given observational dataset, so the experimenter can subsequently design a controlled experiment where the relevant variables are studied further. For this purpose, it is most important that claimed causal relationships are very likely to also be present in the true causal model, while it is less crucial that estimated absent causal relationships reflect the true model. We return to simulation-based evidence of these claims in Section \ref{sec.simstudy.pcges}. 

In conclusion, for this purpose of observational science causal discovery, the error-tradeoff of PC and GES may not be very useful; these algorithms will provide models where effects that are in fact confounded appear unconfounded. This can result in analysts choosing inappropriate analytical tools or designs and, as an effect, obtain faulty causal conclusions along with biased causal effect estimates. Instead, a conservative causal discovery procedure is needed.

\section{Supervised machine learning for causal discovery}
\label{sec.sldisco}

We here propose a supervised machine learning approach for conducting causal discovery, and we will refer to this method as \textit{supervised learning discovery} (SLdisco). SLdisco is motivated by the three limitations of existing causal discovery algorithms highlighted above. 
The main ideas of SLdisco is to 1) avoid poor small sample performance due to statistical error propagation by learning the full causal structure jointly, 2) avoid a preference towards sparse (or dense) graphs by building a machine learning model that is agnostic towards outputted sparsity, and 3) obtain a direct means to affect error trade offs by outputting probabilities of edge/orientation presence rather than binary decisions, whose error trade offs can then be controlled using thresholding and post-processing. SLdisco consists of three steps (further details below):

\begin{enumerate}
\item Simulate training data with known data generating mechanisms.
\item Train a machine learning model on the simulated training data using true CPDAGs as labels.
\item Use the resulting classification function as a one-step causal discovery procedure on real data.
\end{enumerate}

We use the following notation: The terms variable and node are used interchangeably. We use $p$ to denote the number of nodes in the DAG/CPDAG. We denote the $p$ variables by $X_1, ..., X_p$, and we use $n$ to denote sample sizes for each simulated dataset, while $b_\text{train}$ and $b_\text{test}$ are used to denote the size of the training data and testing data, respectively.

\subsection{Data simulation}

We limit the  scope of the method to linear Gaussian data generating mechanisms. In this case, the correlation matrix is a sufficient statistic and hence it is meaningful to only consider these as input for the discovery method. 

The data are simulated in a three-step procedure:
\begin{enumerate}
\item Construct DAG with randomly drawn density (0-80\% missing edges compared to fully connected).
\item Simulate from a linear Gaussian structural equation model according to the DAG, and compute the correlation matrix. 
\item Construct CPDAG adjacency matrix corresponding to the DAG.
\end{enumerate}
Technical details about the data simulation are provided in Appendix \ref{appendix.datasim}, and here we only provide a brief overview. 

The DAGs in step 1) are constructed such that the distribution of the number of edges is uniform on the interval $(m(p)_\text{min}, m(p)_\text{max})$ where
$m(p)_\text{max} = \sum_{i = 1}^{p - 1} i$ (fully connected DAG) and $m(p)_\text{min} = m(p)_\text{max} \cdot 0.2$. We thus consider a broad distribution of graph densities and include very dense graphs in the simulation setup.

We simulate each variable $X_i$ according to the DAG (which holds information about parent sets) using a structural equation:
$$ X_i := \sum_{X_j \in \text{pa}(X_i)} X_j \cdot \beta_{j,i} + \epsilon_i$$
where $\epsilon_i \sim N(0, \sigma_i^2)$ independently. All parameter values ($\beta_{j,i}$ and $\sigma_i$) are drawn randomly.

As data features, we use correlation matrices computed on the full dataset of $X_1, ..., X_p$ variables. As data labels, we use the CPDAGs encompassing the true DAGs. We represent the CPDAGs by their adjacency matrices.

\subsection{Machine learning model}

We use neural networks to construct a mapping from correlation matrices to adjacency matrices. In the special case where the DAG is fully identifiable from observational data (e.g., linear Gaussian data generating mechanisms with equal residual variances), this learning task is achievable since 1) DAG discovery can be formulated as a continuous optimization problem \cite{zheng2018,zheng2020} and 2) a neural network is a universal approximator \cite{hornik1989}.
Hence, with oracle correlation matrix information (corresponding to $n \to \infty$), a sufficiently large training data set and a good neural network architecture, it is possible to estimate adjacency matrices at any level of accuracy in this special case. Clearly, oracle correlation matrix information is not available in real-data settings, so in practice, with a sufficiently sophisticated architecture and sufficiently large training data size (which may both be feasible to obtain), we should be able to achieve a level of accuracy that is \textit{only} limited by how well we estimate the correlation matrix (or, more generally, a sufficient statistic). We find it plausible that a similar result holds for our more general case of CPDAG estimation, and we believe that the results provided below provide evidence for this claim. 

In this article, we do not attempt to provide such a sufficiently sophisticated architecture or training data size. Instead, we present results for a simple, albeit useful convolutional neural network with a set choice of training data size ($b_\text{train} = 1,000,000$), in order to provide a proof of concept. Note that appropriate size of the training data pairs will of course depend strongly on the number of nodes in the data generating DAG. 

\subsubsection{Neural network and post-processing}

We use a simple convolutional neural network to learn the relationship between correlation matrices and CPDAG adjacency matrices. Technical details about its architecture and training are provided in Appendix \ref{appendix.nn}. 

Most importantly, for the final output layer, we use a sigmoid activation function (also known as the \textit{expit)} function, i.e. the inverse of the logit function). This means that our neural network model outputs a matrix of probabilities rather than an adjacency matrix. Each element in this probability matrix then corresponds to the estimated probability of that element in the adjacency matrix being one. 

This matrix of probabilities then needs to be post-processed to obtain an adjacency matrix. Let $O$ denoted the probability matrix. We propose and evaluate two alternative approaches for this, both using a threshold $\tau \in (0,1)$:
\begin{description}
\item[\textbf{Cut-off:}] Set adjacency matrix elements to 1 if they are larger than $\tau$, and set all others to zero. 
\item[\textbf{Backwards PC-orientation (BPCO):}] First use the cut-off post processing method and check whether the resulting matrix, $M$, is a proper CPDAG adjacency matrix. If so, return $M$. Otherwise, repeat the following steps until $M$ is a proper CPDAG adjacency matrix:
\begin{enumerate}
	\item Among the remaining non-zero elements in $M$, choose the element that had the lowest probability in $O$ and set this element to zero. If $M$ is now a proper CPDAG, return it. 
	\item Otherwise, create a new adjacency matrix $\tilde{M}$ by modifying $M$ such that only the skeleton and $v$-structures are preserved. Then apply the orientation rules from the PC algorithm \cite{meek1995} to $\tilde{M}$ and check whether the resulting matrix is a proper CPDAG. 
			\begin{enumerate}
				\item If so, return it.  
				\item If not, discard $\tilde{M}$ and go to step 1. 
			\end{enumerate}
\end{enumerate}
\end{description}

Note that BPCO will always produce a proper CPDAG: If $M$ ends up with only one edge, $\tilde{M}$ will only have a single undirected edge, which is a proper CPDAG. Moreover, BPCO will always either have the same skeleton as the cutoff method or a sparser one, since the procedure in step 2 only changes edge orientations, not adjacencies.

\section{Metrics}
\label{sec.metrics}

As discussed in Section \ref{sec.errortradeoff}, we aim for a causal discovery procedure that is conservative, and our choice of relevant metrics will of course reflect this. We will divide the question into two parts: 1) Adjacency performance and 2) orientation performance. In both cases, we report mean metrics both over the full test data, as well as stratified by the true graph density. Moreover, we also consider the estimated number of edges, and compare this to the true number of edges, in order to assess which threshold level ($\tau$) obtains the closest approximation to the true number of edges. 

\subsection{Adjacency metrics}
Adjacency performance only involves the skeleton of the CPDAG, and as mentioned above, we wish to estimate a supergraph of that skeleton to achieve a conservative causal model estimate. This motivates our first metric of interest, namely the negative predictive value (NPV). Let $\text{TP}, \text{TN}, \text{FP}, \text{FN}$ denote the number of true positive, true negative, false positive and false negative classifications (in this case: adjacencies), respectively. Then the NPV is defined as
$$ \text{NPV} = \frac{\text{TN}}{\text{TN} + \text{FN}}$$
i.e., the probability of an estimated missing adjacency to also be missing in the true skeleton. A high value of NPV will thus reflect a conservative skeleton estimate: Missing adjacencies are generally trustworthy. However, one could of course obtain a NPV of 1 simply by trivially returning a fully connected skeleton (using the convention that $\text{NPV} = 1$ for this degenerate case where there are no classified negatives). Thus we also need to consider an secondary metric that describes the error for positive findings, i.e. adjacencies estimated to be present.

For this purpose, we will use the F1 score:
$$\text{F1} = 2 \cdot \frac{\text{precision} \cdot \text{recall}}{\text{precision} + \text{recall}}.$$
F1 describes a tradeoff between two metrics, precision and recall, where
$$\text{precision} = \frac{\text{TP}}{\text{TP} + \text{FP}} \; \text{ and } \; \text{recall} = \frac{\text{TP}}{\text{TP} + \text{FN}}$$
F1 thus summarizes errors related to positive findings: Precision (also known as the positive predictive value) measures the probability of an estimated adjacency to be a true adjacency, while recall (also known as sensitivity) measures the probability of the estimated graph including a given true adjacency. 

In order to obtain a conservative \textit{and} informative model we thus aim for 1) a large NPV (to obtain a supergraph of the skeleton) and 2) an acceptable F1 (so that the supergraph is close to the true skeleton).

\subsection{Orientation metrics}
We measure orientation performance by use of conditional endpoint metrics, that is, for each correctly identified adjacency, we consider whether each edge endpoint is correctly estimated. A correctly placed arrowhead is considered a true positive, a correctly placed tail is a true negative, a falsely placed arrowhead is a false positive, while a falsely placed tail is a false negative. 

As discussed in Section \ref{sec.errortradeoff}, a conservative causal model should rather have too many undirected edges than too many falsely directed edges. Hence, for orientation, a conservative procedure should prioritize obtaining a high value of precision (i.e. positive predictive value). With a high precision, estimated arrowheads will generally be trustworthy, while estimated tails may or may not reflect the true graph. As for the adjacency metrics, a trivially large value of precision can of course be obtained by simply not orienting any edges. Hence, we again need a secondary metric that summarizes errors related to the opposite (now: false) classifications. 

To this end, we propose a negative classification equivalent to F1: 
$$G1 =  2 \cdot \frac{\text{NPV} \cdot \text{specificity}}{\text{NPV} + \text{specificity}}$$
where 
$$\text{specificity} = \frac{\text{TN}}{\text{TN} + \text{FP}}$$
This metric measures the tradeoff between two types of negative classification errors, just as F1 does for positive classification errors. 

A conservative causal model should hence have 1) a large orientation precision (thus only claiming orientations that have a high probability of being true), and 2) an acceptable G1 (so that some edges are forced to be oriented).

\section{Simulation study}
\label{sec.simstudy}

We evaluate SLdisco by comparing estimated (pseudo) adjacency matrices with true adjacency matrices on atest data set. We vary the following aspect of the procedure and compare the resulting performances:
\begin{itemize}
\item The number of observations in each simulated training dataset, $n_\text{train}$. We consider \\$n_\text{train} \in \{50, 100, 500, 1000, 5000, 10000, 50000\}$.
\item The number of variables in the data generating mechanism, $p$. We consider $p \in \{5, 10, 20\}$. 
\item The post processing procedure for converting probability matrices to (pseudo) adjacency matrices. We consider the cutoff and BPCO methods. 
\item The threshold value, $\tau$. We consider $\tau \in \{0.01, 0.05, 0.1, 0.2, 0.3, 0.4, 0.5\}$.  
\end{itemize}
We consider all combinations of the above. 

We first compare PC and GES in order to choose a single benchmark causal discovery method among the two, which simplifies subsequent comparisons with SLdisco. 
For PC, we consider a sequence of test significance levels, namely $\alpha \in \{10^{-8}, 10^{-4}, 10^{-3}, 0.01, 0.05, 0.1, 0.2, 0.5, 0.8\}$. For GES, we use the Bayesian information criterion (BIC) to score models. In Appendix \ref{appendix.results.extrafig}, we furthermore compare this choice to an alternative score, namely a modified version of BIC with a larger penalty for the number of variables. 

For each evaluation scenario, we train the neural network on $b_\text{train} = 1,000,000$ correlation matrix/adjacency matrix pairs, and we  test the procedure on $b_\text{test} = 5000$ correlation matrix/adjacency matrix pairs. We report performance metrics averaged over these 5000 repititions for each scenario. 

Data simulations and computations for SLdisco and PC are conducted in R. We use the \texttt{keras} package \cite{allaire2021} for neural network training, the \texttt{pcalg} package \cite{kalisch2012} for PC computations, and the \texttt{causalDisco} \cite{petersen2021-b} package for evaluations. GES computations are performed in TETRAD \cite{ramsey2018} using the fast GES (FGES) version of GES \cite{ramsey2017}. All code, as well as the simulated data, is available online at \url{https://github.com/annennenne/SLdisco}.

\subsection{PC and GES comparison}
\label{sec.simstudy.pcges}

\begin{figure}
\centering
\includegraphics[width = \textwidth]{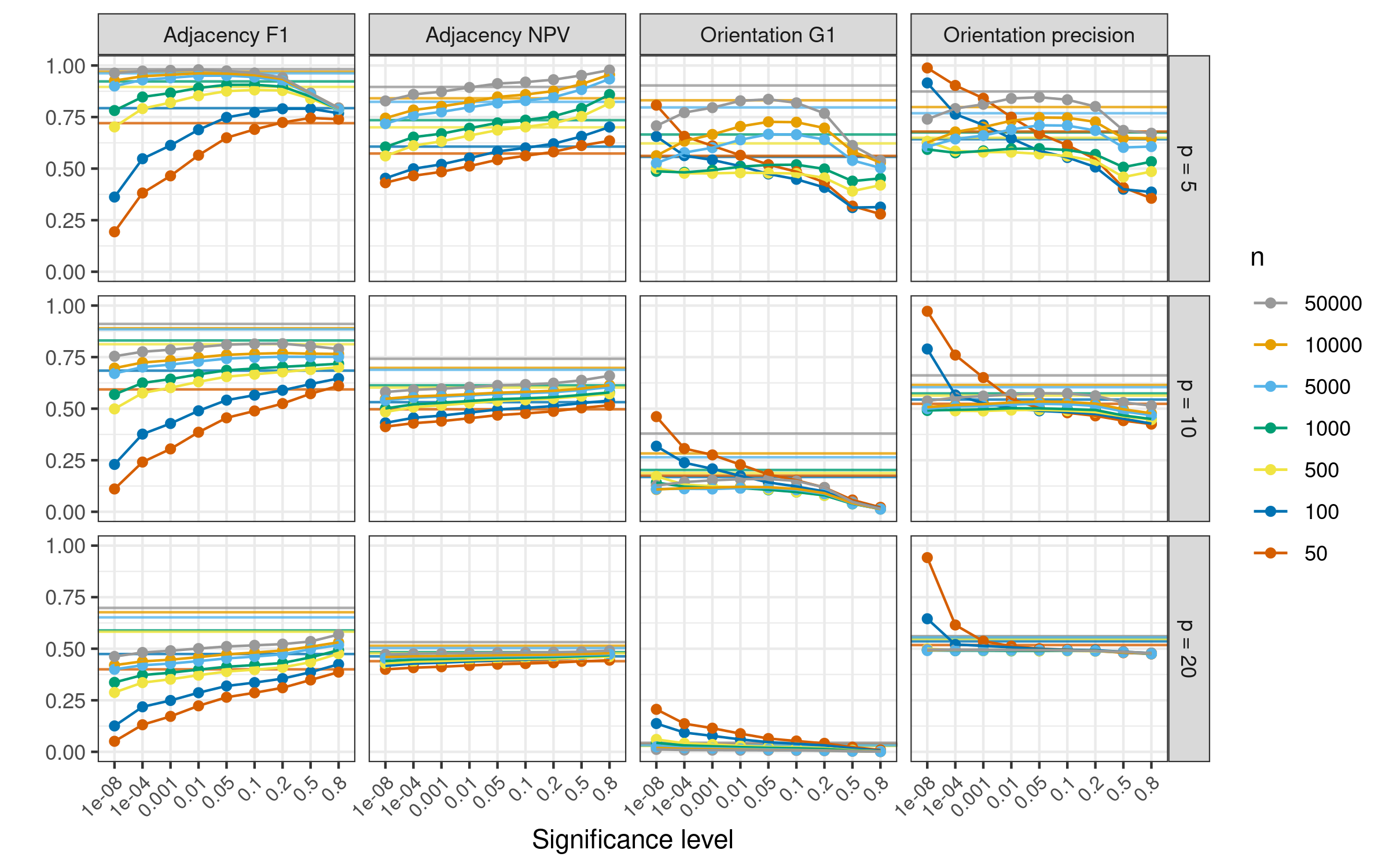}
\caption{All metrics for PC (lines with dots) for varying values of significance level, compared to GES with BIC scoring (lighter straight lines).}
\label{fig.pc.ges.all}
\end{figure}

Figure \ref{fig.pc.ges.all} compares PC and GES with BIC scoring. For the small $p = 5$ graphs, we see that it is possible to choose a significance level for PC (around $0.05$) so that the two methods perform equally well in terms of adjacency F1, adjacency NPV and orientation precision when $n \geq 500$. However, for $p \in \{10, 20\}$, GES consistently produces better performance for the adjacency metrics, no matter the choice of significance level for PC. For the orientation metrics, GES performs better than PC in most cases, and in the cases where PC is superior, both methods perform very poorly (for example, orientation G1 for $p = 20$). All in all, we thus find GES to be the strongest competitor. 

In Appendix \ref{appendix.results.extrafig}, we compare two different versions of GES using different choices of score and find that standard BIC performs best overall (see Figure \ref{fig.ges.bothscores}). In the following, we will therefore use GES with BIC score as our main benchmark for SLdisco, and not include results from PC nor other choices of GES score.

\subsection{Estimated number of edges}
\label{sec.noedges}
\begin{figure}
\centering
\includegraphics[width = \textwidth]{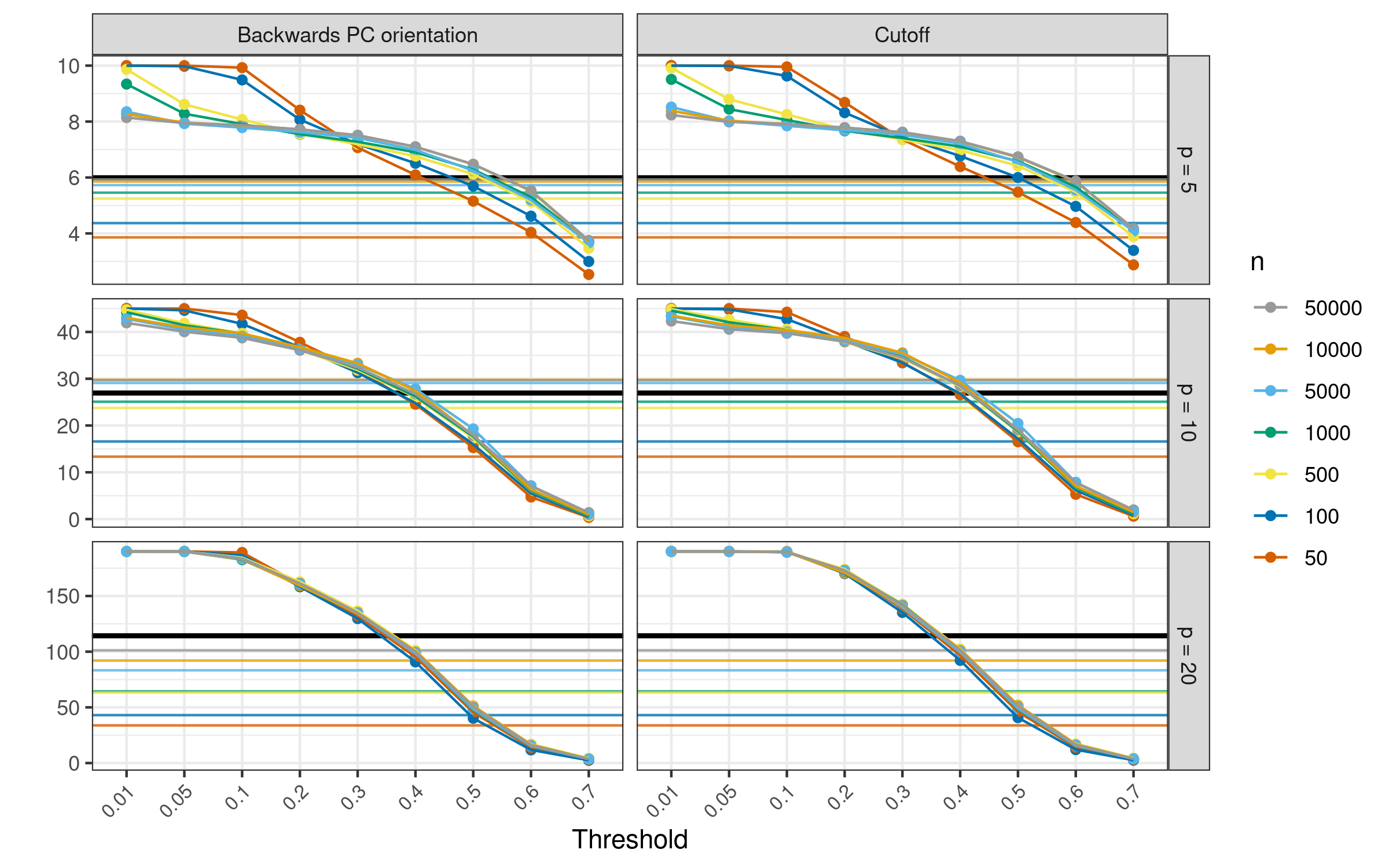}
\caption{Estimated and true numbers of edges, stratified by number of nodes ($p$), sample size ($n$) and post-processing method. The colored straight lines present the average estimated number of edges for GES, while the colored lines annotated with points present the average estimated number of edges by SLdisco. The black, straight line marks the average true number of edges for each value of $p$.}
\label{fig.nedges}
\end{figure}

Figure \ref{fig.nedges} presents the number of estimated edges for SLdisco and GES. We find that GES generally produces CPDAGs with too few edges (except for the largest sample sizes when $p = 10$). SLdisco is able to obtain an average number of edges that is close to the truth by setting the threshold appropriately. We find the closest approximation of the true number of edges at $\tau = 0.4$ for $p = 5$, at $\tau = 0.4$ for $p = 10$, and at $\tau = 0.3$ for $p = 20$. We propose that in practice, $\tau$ may be selected by referring to these results.

We also find a rather good approximation of the correct number of edges when stratifying by quartiles of true numbers of edges (see Figure \ref{fig.nedges.bytruth} in Appendix \ref{appendix.results.extrafig}), though SLdisco underestimates the number of edges for sparse graphs when $p = 20$.

\subsection{Adjacency results}
\begin{figure}
\centering
\includegraphics[width = \textwidth]{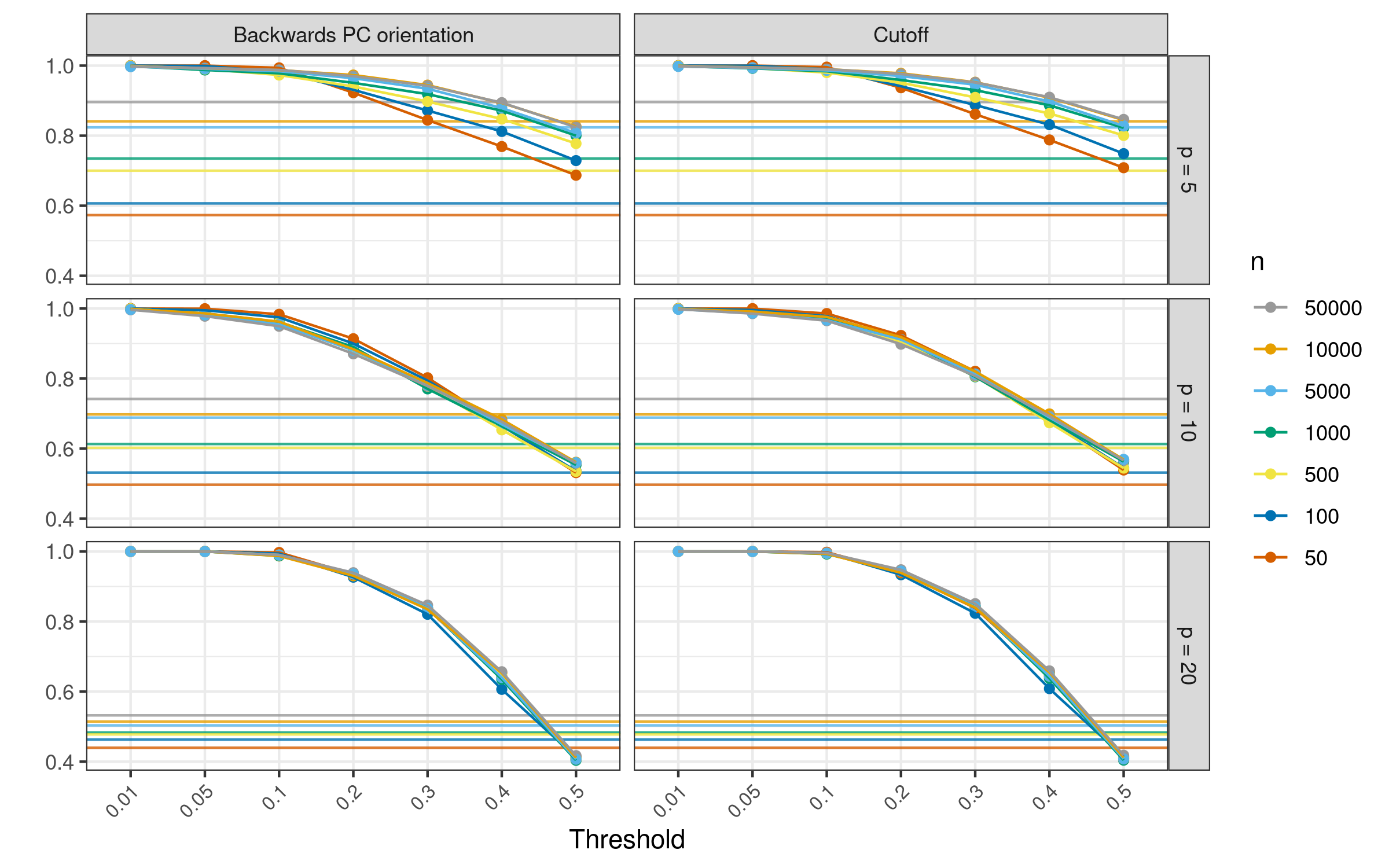}
\caption{Adjacency negative predictive values for SLdisco (lines with points) and GES (lighter straight lines). The results are stratified by number of nodes ($p$), sample size ($n$) and post-processing method.}
\label{fig.adj.npv}
\end{figure}

\begin{figure}
\centering
\includegraphics[width = \textwidth]{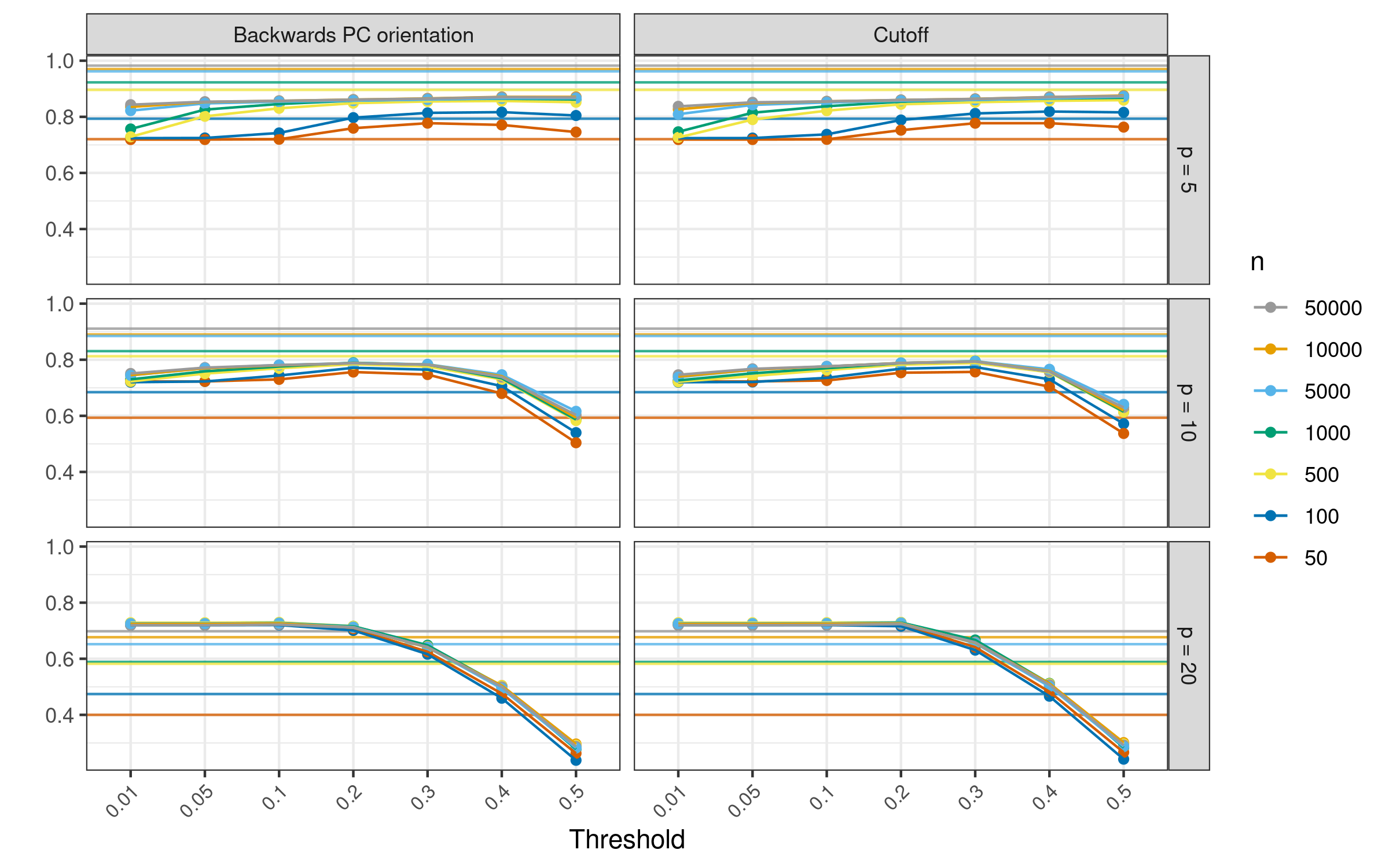}
\caption{Adjacency F1 scores for SLdisco (lines with points) and GES (lighter straight lines). The results are stratified by number of nodes ($p$), sample size ($n$) and post-processing method.}\label{fig.adj.f1}
\end{figure}

Figure \ref{fig.adj.npv} shows adjacency NPV. For the cutoff method, we see that for all values of $p$, when $\tau \leq 0.3$, SLdisco produces a NPV that is better than that of GES. Especially for larger values of $p$, we see a marked improvement in NPV, compared to GES. The BPCO post-processing method produces similar results, although generally with slightly lower NPV values. Both SLdisco methods show little sensitivity towards $n$, compared with GES. 

Figure \ref{fig.adj.f1} presents adjacency F1 scores. We again find similar results for the cutoff and BPCO methods, and we therefore report results only for the cutoff method. For $p \in \{5, 10\}$, we see that SLdisco is only able to achieve similar or larger F1 scores than GES for the smallest sample sizes, while GES produces larger F1 scores for $n \geq 500$. This does not depend on the choice of threshold for SLdisco. For $p = 20$, SLdisco provides larger or equal values of F1 than GES when $\tau \leq 0.2$. We see little sensitivity to sample size for SLdisco, while GES performance varies quite a bit depending on $n$.

\subsection{Orientation results}
 
\begin{figure}
\centering
\includegraphics[width = \textwidth]{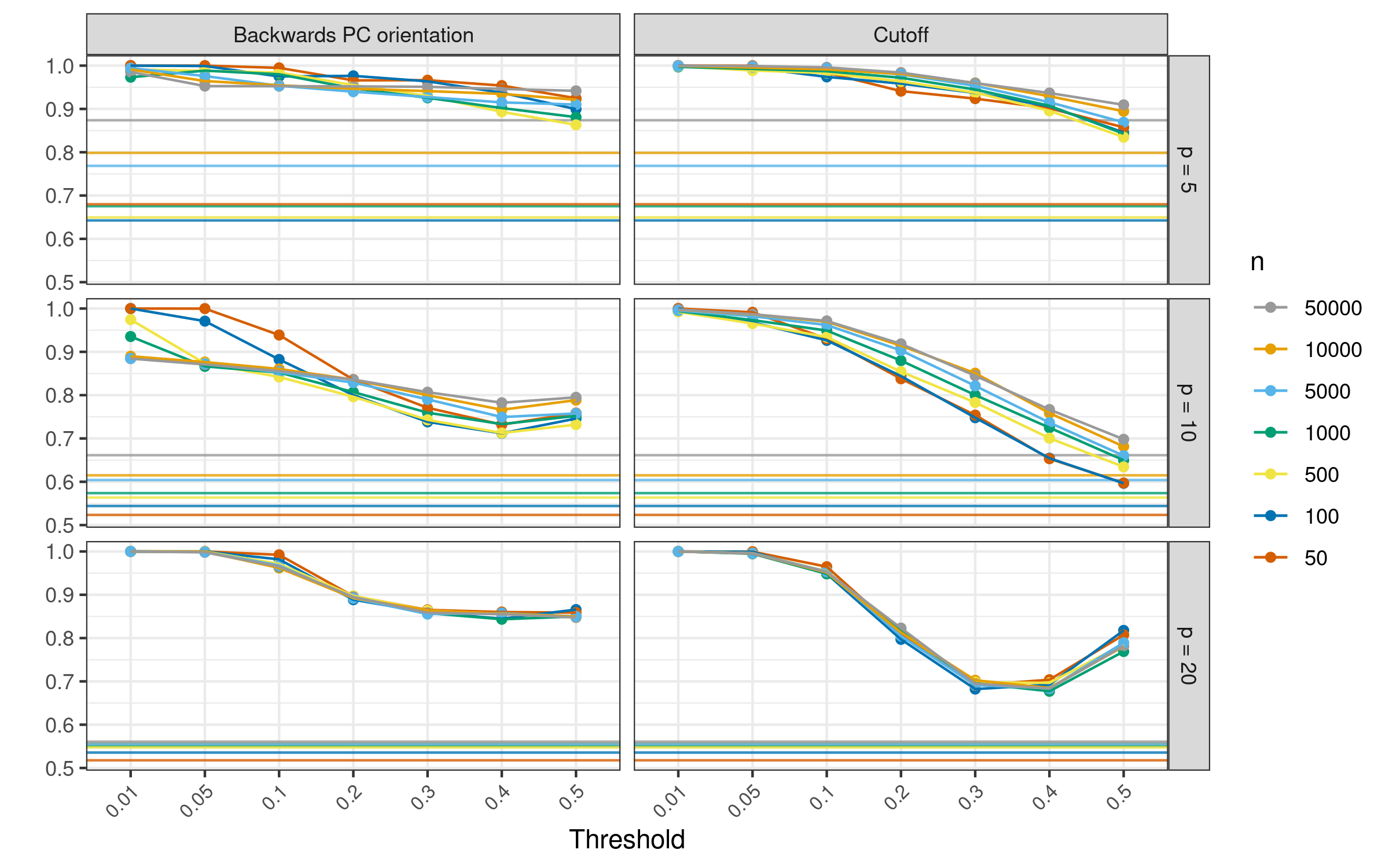}
\caption{Conditional orientation precision values for SLdisco (lines with points) and GES (lighter straight lines). The results are stratified by number of nodes ($p$), sample size ($n$) and post-processing method.}
\label{fig.dir.precision}
\end{figure}

\begin{figure}
\centering
\includegraphics[width = \textwidth]{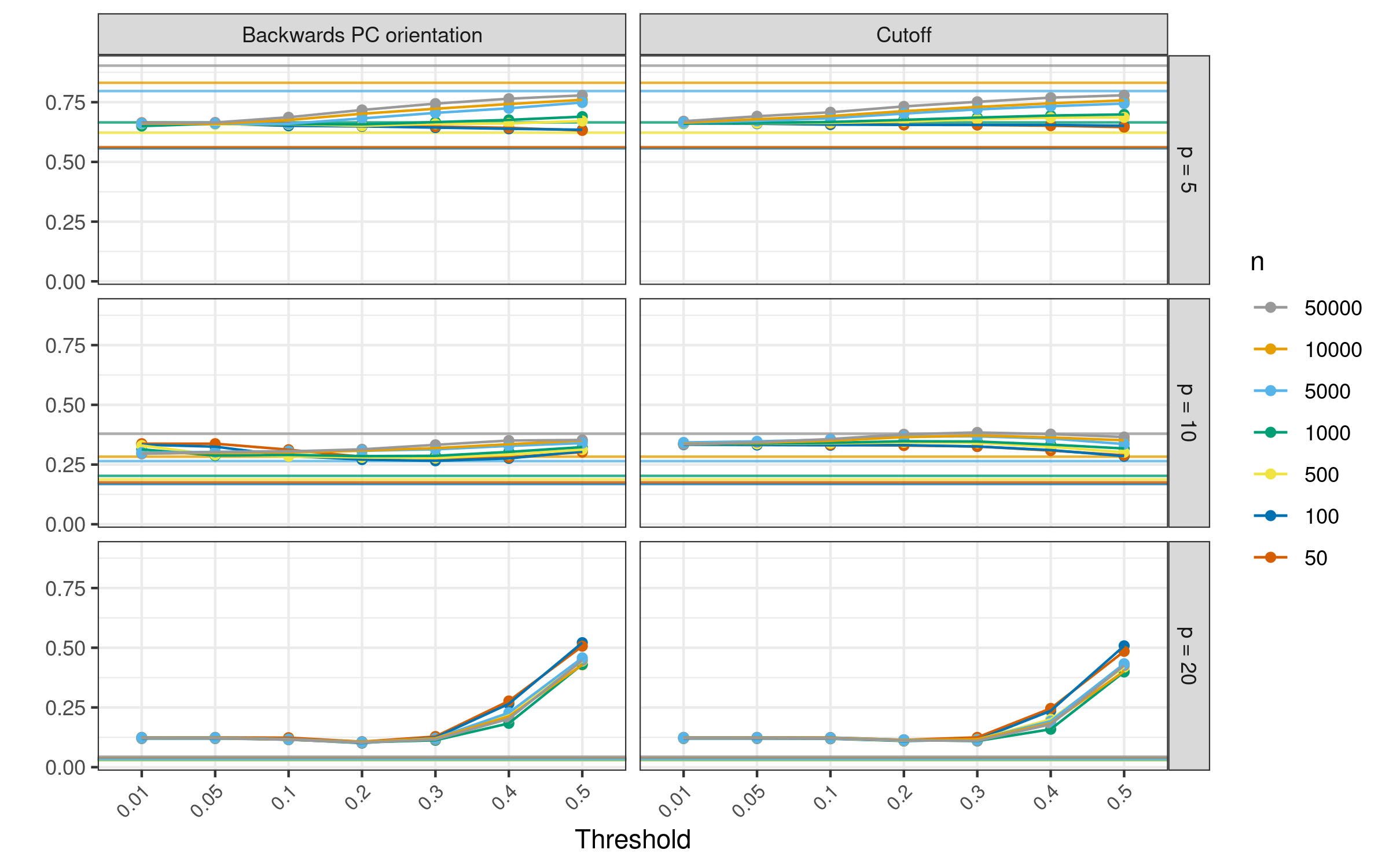}
\caption{Conditional orientation G1 scores for SLdisco (lines with points) and GES (lighter straight lines). The results are stratified by number of nodes ($p$), sample size ($n$) and post-processing method.}\label{fig.dir.g1}
\end{figure}

We now turn to conditional orientation metrics. Figure \ref{fig.dir.precision} presents orientation precision for SLdisco and GES. We find that SLdisco outperforms GES for all combinations of $n$ and $p$, no matter the choice of $\tau$. 

Figure \ref{fig.dir.g1} shows orientation G1 score for SLdisco and GES. For $p = 20$, we find that SLdisco performs better or equally well as GES for all values of $\tau$ and $n$. For $p = 10$, the same holds true for $n \leq 10000$, while GES outperforms SLdisco in the $n = 50000$ case for some values of $\tau$ when SLdisco is used with the cutoff postprocessing method, and all values of $\tau$ when SLdisco is used with BPCO. For $p = 5$, GES outperforms SLdisco when $n \geq 5000$. 

For both orientation metrics we once again find SLdisco to be less sensitive towards sample size than GES.

\subsection{Results stratified by true graph density}

\begin{figure}
\centering
\includegraphics[width = \textwidth]{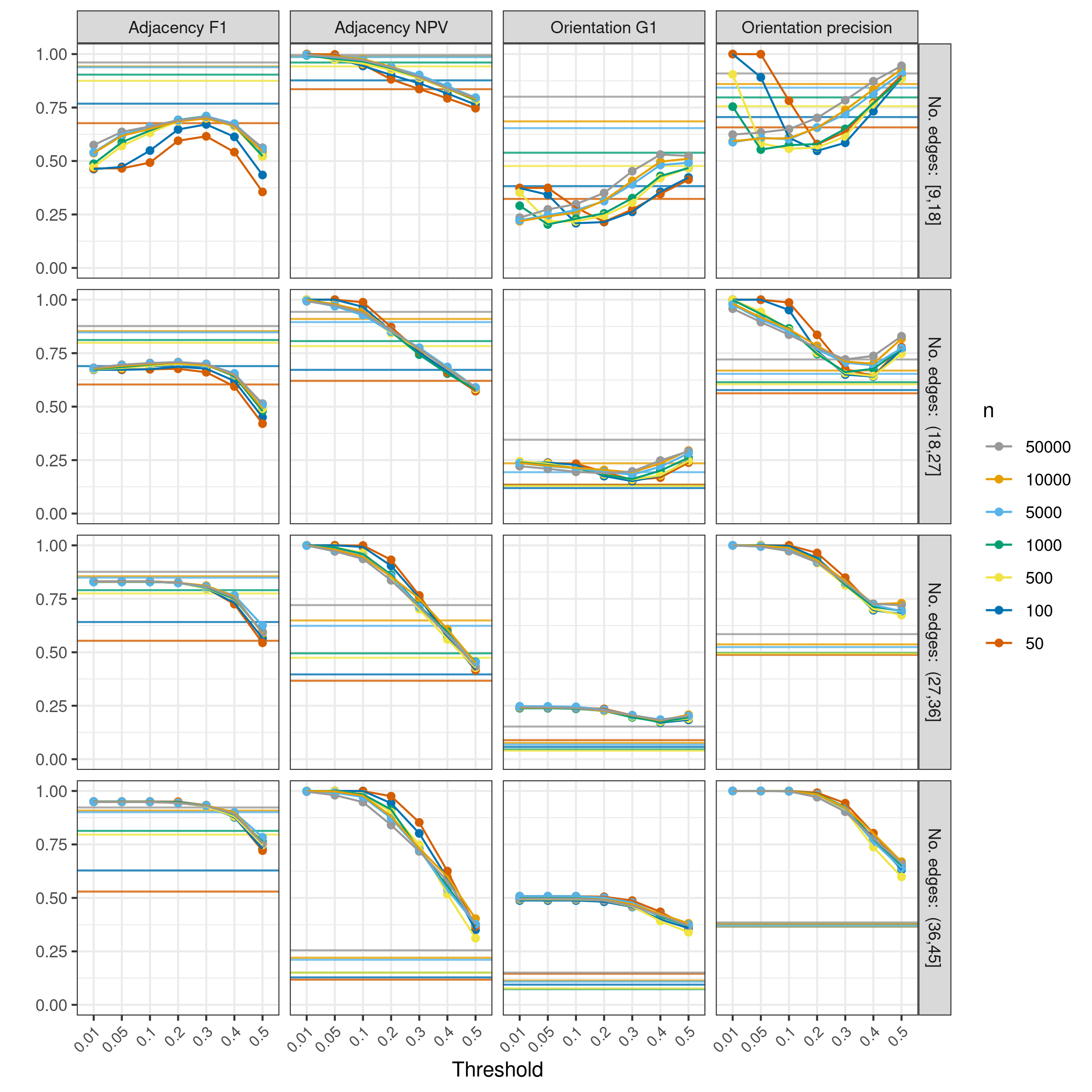}
\caption{Results for SLdisco (lines with points) and GES (lighter straight lines), stratified by true graph edge number quartiles and sample size. Results are for graphs with $p = 10$, and uses BPCO post-processing for SLdisco.}\label{fig.allmet.bynedge}
\end{figure}

Figure \ref{fig.allmet.bynedge} focuses on the BPCO post-processing method. Here, the four metrics from above are stratified by graph density, measured as quartiles of the true number of edges. We only present results for $p = 10$, but find similar results for $p \in \{5, 20\}$. While GES struggles increasingly with adjacency NPV, orientation G1 and orientation precision as the graph density increases, SLdisco
is strongest on relatively dense graphs for these metrics. For the dense graphs, SLdisco is hence able to outperform GES for most sample sizes.

But on the sparsest graphs, SLdisco struggle especially with orientation G1 and adjacency F1. In contrast, GES adjacency F1 performance is less affected by graph density. As in the above, we again see that GES depends more strongly on sample size, while SLdisco is able to produce more similar results for both small and large samples.

\section{Application}
\label{sec.appli}

For the real data application, we apply PC, GES and SLdisco to data from the Metropolit Cohort \cite{osler2004,osler2006}. This is a longitudinal dataset following a cohort of Danish men from their birth in 1953 with data collections in childhood and adulthood, as well as register-based information and followup during other periods of their lives. This is a high-quality dataset with good validity and low degree of measurement error \cite{osler2006}. 

We use a subset of 10 variables also used in \cite{petersen2021}, which includes follow-up until 2018. The original dataset also contains binary variables, and we use only numeric variables for a more direct comparison with the simulation study. One numeric variable (\textit{total years of smoking}) is dropped so that we obtain exactly 10 variables, which means that we can easily apply the pretrained SLdisco models from the simulation study. 

The 10 variables are all assigned to one of four periods: Birth, childhood, young adulthood, or adulthood. We use only complete cases and condition of cohort members being alive and residing in Denmark in 2018. This results in a dataset of $n = 2928$ observations. Appendix \ref{appendix.application.extrafig} provides information about marginal distributions and pairwise associations in this data.

\subsection{Methods}

We of course do not know the true causal model for this data, so we cannot readily apply the same evaluation strategy as in Section \ref{sec.simstudy}. Instead, we will: 
\begin{enumerate}
\item Discuss plausibility of the causal models proposed by SLdisco, GES and PC, respectively. 
\item Evaluate how the performance of these methods varies with sample size. 
\end{enumerate}

For each causal discovery method, we use the settings that provided the closest estimation of the true number of edges in the simulation study for $p = 10$ and $n = 5000$ (the closest value we have considered to $n_\text{metropolit} = 2928$). We use PC with a significance level of $\alpha = 0.1$, GES with BIC scoring and SLdisco with BPCO post-processing and $\tau = 0.4$. 

To address plausibility (task 1), we apply PC, GES and SLdisco to correlation matrices computed on the full Metropolit dataset. We use an SLdisco model trained on datasets with $n = 5000$. We make use of the fact that the Metropolit dataset includes temporal information; each variable is assigned to a specific period and this provides us with a ground truth about \textit{some} of the potential causal orientations: Any edge oriented against the direction of time is necessarily false.

To compare the performance across varying sample sizes (task 2), we use random subsampling. Even though we do not know the ground truth causal model, the more data we have, the better we should be able to estimate it. Hence, we can use CPDAGs estimated from the full Metropolit dataset as a "best estimate", and compare these graphs to graphs estimated on from correlations matrices computed on random subsets of the Metropolit data. This will allow us to evaluate the impact of small sample sizes on real data. We draw a random subsample for each $n \in \{50, 100, 500, 1000\}$, compute the corresponding correlation matrices, and use these as input for PC, GES and SLdisco. The resulting causal models are then compared with the same method's "best estimate", as obtained from task 1. We report the comparisons using the same four metrics as in the simulation study. 

We use the \texttt{pcalg} \cite{kalisch2012} R package for the PC and GES computations, and the \texttt{causalDisco} R package \cite{petersen2021-b} for evaluation and plotting. All code, as well as correlation matrices, is available online at
\begin{center}\url{https://github.com/annennenne/SLdisco}\end{center}

\subsection{Results}

\begin{figure}
\centering
\subfloat[SLdisco \label{subfig.appli.sldisco}]{
\includegraphics[width=0.5\textwidth,trim={1.5cm 1.5cm 0.5cm 1.5cm}]{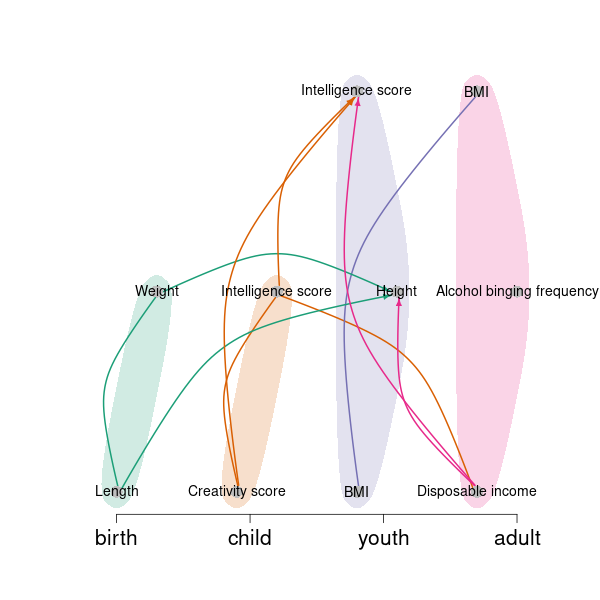}
} 
\subfloat[PC \label{subfig.appli.pc}]{
\includegraphics[width=0.5\textwidth,trim={0cm 1.5cm 0.5cm 1.5cm}]{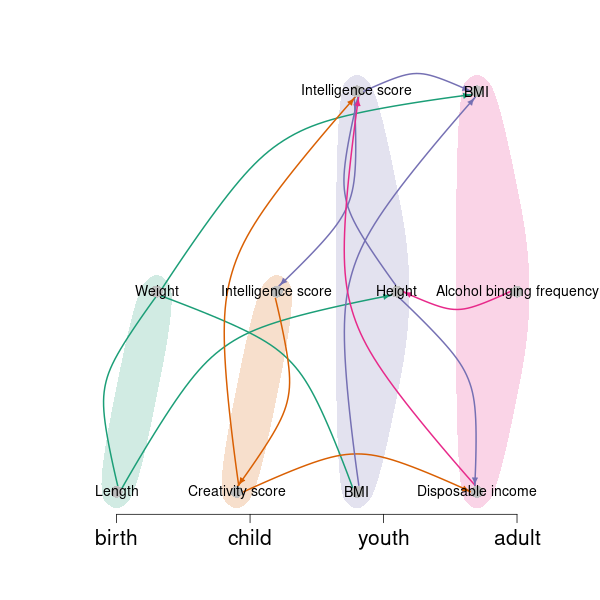}
} \\
\subfloat[GES \label{subfig.appli.ges}]{
\includegraphics[width=0.5\textwidth,trim={1.5cm 1.5cm 0.5cm 1cm}]{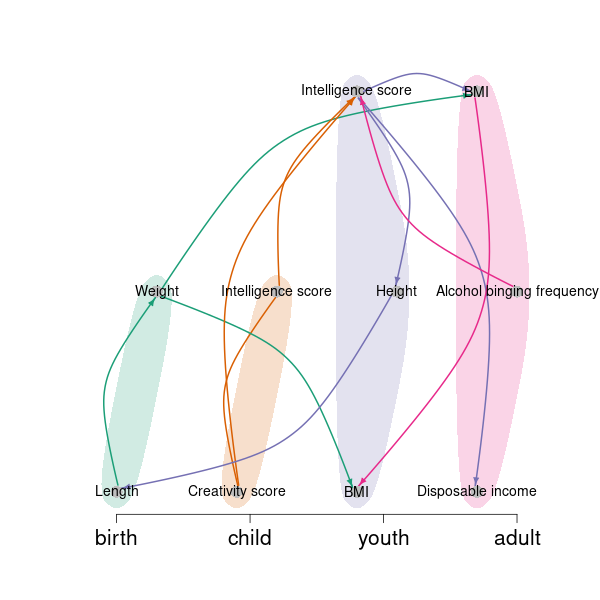}
}
\caption{CPDAG estimates for the Metropolit example based on all $n = 2928$ observations using SLdisco (a), PC (b) and GES (c).}
\label{fig.appli}
\end{figure}

Figure \ref{fig.appli} shows the estimated CPDAGs for the Metropolit data for SLdisco, PC and GES, respectively. The variables are arranged in temporal periods (marked in colors) and edges are colored according to the period the originate from, if they are oriented, or simply their first period, if they are unoriented. None of the three methods are provided with external information about the temporal order of the variables. 

\subsubsection{Plausibility of the estimated models}
For SLdisco we see that a number of plausible causal links are identified: edges between birth weight, birth length and height in youth, edges between creativity score in childhood and intelligence score in childhood and youth, and an edge between body mass index (BMI) in youth and BMI in adulthood. But the CPDAG also includes two edges that were erroneously oriented against the direction of time: An edge from disposable income in adulthood to height in youth, as well as an edge from disposable income in adulthood to intelligence score in youth. The CPDAG also includes an undirected edge between intelligence score in childhood and disposable income in adulthood, which may or may not be plausible. On the one hand, it would be more obvious that a specific effect of intelligence should go via intelligence score in youth. But on the other hand, socio-economic status may affect intelligence test performance in children \cite{vonstumm2015}, and hence the edge between intelligence score in childhood and adulthood income could represent an indirect effect due to lack of social mobility.  

PC also suggest the wrongly oriented edge between disposable income in adulthood and intelligence score in youth, as well as two other erroneously oriented edges: One from alcohol binging frequency in adulthood to height in youth and one from intelligence score in youth to intelligence score in childhood. PC furthermore recovers a number of plausible causal links, e.g., an effect of BMI in youth on BMI in adulthood, and an effect of birth length on height in youth. However, the CPDAG also shows a number of less plausible causal links, for example a causal link between height (youth) and alcohol binging (adulthood), height (youth) and income (adulthood), and a direct effect of birth weight on BMI in adulthood (on top of the more plausible effect via BMI in youth). Furthermore, we note that in the current application, PC does not return a proper CPDAG; the graph in Figure \ref{fig.appli} (b) includes a cycle: 
$\textit{Childhood creativity score} \to \textit{youth intelligence score} \to  \textit{childhood intelligence score} \to \textit{childhood creativity score}$. 
PC may produce graphs that are not proper CPDAGs if there there is conflicting information with respect to separating sets, which can occur if some of the algorithm's assumptions (no unobserved confounding, faithfulness and a correct statistical test of independence) are not fulfilled \cite{kalisch2012}. We return to this point below. 

GES also finds three edges against the direction of time (BMI in adulthood to BMI in youth, alcohol binging in adulthood to intelligence score in youth, and height in youth to birth length), and plausible causal links between the variables related to intelligence and creativity scores. It shares a number of plausible causal links with SLdisco and PC, e.g., links between intelligence score and creativity score variables, as well as links between birth weight and birth length. Among less plausible causal links is the suggested effect of birth weight on adult BMI (not mediated through BMI in youth), and the edge between intelligence score and height, both in youth. 

It should be noted that all three methods rest on a strong assumption that may not be reasonable for this example, namely that there are no unobserved confounders or selection variables. If the example does suffer from such unobserved variables it may induce spurious edges in the outputs of PC \cite{spirtes2000} and GES \cite{ogarrio2016}, which may explain why these methods -- somewhat surprisingly -- produce denser CPDAGs than SLdisco in this application. The fact that PC produces a graph with a cycle also lends evidence to this explanation, as this can be expected behavior when applying PC to data with unobserved confounding \cite{kalisch2012}. We do not know how density estimation for SLdisco is affected by presence of unobserved confounding, and leave this question to future research. Unobserved confounding can also lead to wrong edge orientations for both PC and GES \cite{spirtes2000,ogarrio2016}, and hence it may explain some of the faulty oriented edges that we found above. 

Moreover, all three methods rely on the data being jointly Gaussian, and this assumptions may not be fully satisfied for some variables, especially the three adulthood variables which have unimodal but skewed distributions (see Figure \ref{fig.appli.plotmatrix} in Appendix \ref{appendix.application.extrafig}). All in all, the results presented here should thus be interpreted with some caution, and more work should be dedicated to further investigating the impact of the strong assumptions that SLdisco relies on, and whether they may be relaxed. 

\subsubsection{Influence of sample size}
\begin{table}
\centering
\begin{tabular}{l | r | r | r | r | r }

\textbf{Method} & \textbf{Subsample} $\mathbf{n}$ & \textbf{Adj. F1} & \textbf{Adj. NPV} & \textbf{Ori. G1} & \textbf{Ori. precision} \\
\hline 
SLdisco & 50 &  0.67 & 0.88 & 0.75 & 1.00 \\
& 100 & 0.67 & 0.88 & 0.75 & 1.00 \\
& 500 & 0.89 & 0.95 & 0.55 & 1.00 \\
& 1000 & 0.95 & 0.97 & 0.80 & 1.00  \\
\hline
PC & 50 & 0.53 & 0.78 & 0.33 & 1.00 \\ 
& 100 & 0.53 & 0.78 & 0.33 & 1.00 \\
& 500 & 0.72 & 0.85 & 0.20 & 0.50 \\
& 1000 & 0.75 & 0.86 & 0.33 & 0.71 \\
\hline
GES & 50 & 0.56 & 0.82 & 0.33 & 1.00 \\ 
& 100 & 0.67 & 0.85 & 0.29 & 1.00 \\
& 500 & 0.64 & 0.86 & 0.00 & 1.00 \\
& 1000 & 0.76 & 0.89 & 0.00 & 0.25 
\end{tabular}
\caption{Results for subsampled Metropolit data applications. For each method, the subsample based results are compared to results from the same method based on the full Metropolit dataset ($n = 2928$).}
\label{tab.appli}
\end{table}

Table \ref{tab.appli} presents results from the subsampling experiment. We see that SLdisco produces very similar CPDAGs across the different values of $n$, as reflected by generally large metric values. The large values of the adjacency metrics show that the estimated graph skeletons are quite similar, and increasingly so when $n$ increases. Unoriented edges are preserved across subsamples (as evident from the orientation precisions), whereas estimation of unoriented edges differs a bit more. In comparison, GES produces lower values and hence is more sensitive towards sample size, and PC is even more sensitive than GES.

\section{Discussion}
\label{sec.discussion}

We have proposed a new method for causal discovery based on supervised machine learning, SLdisco, and we have evaluated it focusing on the three highlighted challenges for using existing causal discovery methods in observational sciences, namely 1) poor small sample performance, 2) bias towards sparse graphs, and 3) error-tradeoffs that result in non-conservative causal model estimates.

In a simulation study, we found that SLdisco produces more conservative causal models than both PC and GES: SLdisco generally achieves larger adjacency NPVs and orientation precisions than PC and GES, which means that adjacency absence and oriented edges proposed by SLdisco are more trustworthy. These increases in conservativeness come at the price of modestly reduced informativeness (as measured through adjacency F1 and orientation G1); SLdisco is overall not as good at detecting adjacency presence and orients fewer edges than PC and GES.  

Moreover, we found that SLdisco shows promising potential for estimating the correct graph density. We also find that SLdisco is more robust towards graph density and that it shows little sensitivity towards sample size.

In the application, we found SLdisco to produce a mostly plausible CPDAG, whereas PC and GES identified more implausible causal links or orientations. However, their results were also denser, and this suggests that the strong assumptions underlying all three methods -- Gaussianity and absence of unobserved confounding  -- may not be fulfilled for this application, and hence further real data assessments are needed. We were however able to systematically assess the influence of sample size on the estimated causal models in the application as well, and we found that SLdisco was again much less sensitive towards sample size than the other two methods. SLdisco thus makes better use of information available in small sample data, both on simulated and real life data. All in all, we conclude that SLdisco holds the potential to address the three challenges for causal discovery in observational sciences. 

We have compared SLdisco with two well-known and well-studied procedures for causal discovery, namely PC and GES, and we have considered a simplified setting with Gaussian data and no unobserved confounding. This is obviously a limitation of the study in terms of generalizability, as we will return to below. However, it is also a strength in terms of validity of the conclusions. PC and GES are correct and complete for the type of data studied here, and hence we are able to directly study the finite-sample statistical properties of the methods without blurring the picture by other factors. The only reason for any deviance from fully correct estimated CPDAGs are necessarily due to finite sample properties of the methods. By comparing SLdisco to these two methods, and by focusing on Gaussian data, we thus get a good picture of how sequential testing (for PC) or greedy searching (for GES) impacts finite sample causal discovery performance. 

We now turn to the limitations imposed by the use of Gaussian training data for SLdisco. Gaussianity may not always be a reasonable assumption, and it is not clear to what extent the performance of SLdisco depends on this. It is difficult to conduct a general simulation-based investigation into the practical influence of deviations from Gaussianity: We would have to simulate data under a specific alternative distribution, and would hence run into the very same generalization issues, and still not be much more knowledgeable about the performance on real data, where the data generating mechanism is unknown. The real data application did include some variables whose distributions were quite skewed and hence deviates from Gaussianity. Nonetheless, SLdisco produced a reasonable CPDAG with only two definitely erroneous edges. Both of these edges were problematic due to their orientation, but plausible in terms of their adjacencies. This could suggest that Gaussianity is not a crucial assumption for SLdisco, at least when the variable distributions are unimodal as in our application. Applying SLdisco to more data with varying distributions will be helpful in gaining a better understanding of the influence of the Gaussianity assumption. 

On a related note, our choice to use correlation matrices as input is of course only equivalent with using the full data under the assumption of Gaussianity, where the correlation matrix is a sufficient statistics (if we disregard the data scale). Our suggested procedure could easily be modified to use full datasets, rather than correlation matrices, as its input. Using full datasets as inputs for SLdisco should be more robust towards deviations from Gaussianity and would allow SLdisco to utilize information in the data beyond partial correlations. Using full data would, however, impose a larger burden on computer memory, especially when $n$ is large. 

The assumption of no unobserved confounding, on the other hand, is potentially more crucial. This assumption is very strong and might not be fulfilled in many observational science applications. The CPDAGs proposed by GES and PC in the application may suggest that there are issues due to unobserved confounding in the Metropolit dataset. If so, it is interesting that SLdisco does not seem to be influenced by deviation from this assumption in the same way as PC and GES: It appears that SLdisco does not add extra spurious adjacencies -- like the other two methods are known to do. A possible explanation for this could be that SLdisco is more robust towards unobserved confounding, but further work is needed to assess if this holds. In either case, extending SLdisco to also handle unobserved confounding could be achieved rather easily simply by simulating data in the presence of unobserved confounding. However, causal model equivalence classes would then have to be represented by partial ancestral graphs (PAGs) (instead of CPDAGs) \cite{richardson2002}, which are more complicated graphical objects that include more edge types (bidirected, directed, undetermined, and combinations of the latter two). Expanding SLdisco to also handle data with unobserved confounding would therefore also require some modifications to the output layer of the proposed neural networks, and the orientation metrics used for evaluations would have to be generalized to handle more edge orientation possibilities. 

Another limitation of our suggested machine learning procedure is our proposed neural network architecture. The architecture is clearly simplistic and we expect that performance can be greatly improved by substituting our approach with a more sophisticated and tailored architecture. \cite{li2020} remark that the causal discovery task is \textit{permutation equivariant} in the sense that permuting variable order in the input (i.e. permuting both rows and columns in the correlation matrix) results in the same permutation in the output (permuting both rows and columns in the adjacency matrix). They propose to use a equivariant neural network to natively support this feature, although with a different causal discovery task in mind (DAG discovery). Using such a permutation equivariant architecture in our setting could also be interesting. This would also ensure that the method becomes completely order invariant, which means that different orderings of inputted variables result in the same outputted adjacency matrix. This is an attractive and natural property to request of a causal discovery method, however, for some commonly used approaches, including PC, it is not completely fulfilled \cite{colombo2014}. 

We found that the post-processing method used to convert probability matrices to (pseudo) adjacency matrices was not very influential on the overall performance. The BPCO, which ensures that the output is a proper CPDAG, generally obtained slightly worse performance than the cutoff method for the adjacency metrics, but comparable or better performance in the orientation metrics. 
It may be possible to construct a better post-processing method since there are no guarantees that our proposed greedy method will find an optimal solution from the estimated probability matrix. However, we believe that a more useful line of research would be to natively include CPDAG structure requirements in the neural network optimization, so that no or little post-processing becomes necessary. This could be done by changing the loss function used to optimize the neural network. The current choice of loss function, the binary cross-entropy loss, essentially considers each entry in the output matrix one by one and seeks to optimize it locally, without taking the structure of the full output matrix into account. Informally, this loss is not aware of the fact that it should be looking for a CPDAG. Naturally, if the neural network is sufficiently sophisticated and $n \to \infty$, $b_\text{train} \to \infty$, optimizing the adjacency matrix entry-by-entry will converge to the correct solution. But with finite $n$ and $b_\text{train}$, we suspect much can be gained by choosing a loss function that better captures the estimation goal. However, constructing such a function is not straightforward. Neural network weights are optimized using a variation of gradient descent and hence the loss function needs to be differentiable, and there has to exist analytical expressions for its derivatives, as they are computed a very large number of times. We leave suggestions for candidates for such a loss function to future research. 

A strength of SLdisco is the very low computation time at inference time, that is, when the method is used on real data. A pretrained SLdisco model provides  a mapping from input data to CPDAGs that does not need further optimization. SLdisco will therefore be easy to combine with e.g. stability selection or other bootstrapping procedures to provide more insights into the statistical uncertainty of causal discovery --  a topic that is currently heavily under-described. At training time, SLdisco does require some computation time (approximately 100-200 seconds per epoch resulting in 4-8 hours for the models presented here). However, it is only necessary to train a neural network once for each $n$-$p$ combination. Moreover, transfer learning across similar values of $n$ may be permissible  so that new models need to be trained less often. Note also that more efficient training schemes may be achievable by e.g. training first on a large $n$ and then only fine-tuning weights when considering a different value of $n$. For $p$, such transfer learning is less natural, as we do not expect the mapping from correlation matrices to adjacency matrices to be the same for different values of $p$. This may be tested empirically in future studies, but requires that a given neural network can input correlation matrices of varying sizes. One way to obtain this feature is by simply zero-padding or up-sampling the input matrix in the first layers of the architecture. Alternatively, if for example graph neural networks are used, this feature is ensured natively \cite{scarselli2008}. This neural network type may also be able to avoid two other issues raised here, lack of permutation equivariance and the need of post-processing. We believe that applying graph neural network in SLdisco will be an interesting and natural next step in further improving the method we have proposed here.

\bibliography{nndisco-biblo}

\newpage

\appendix

\section{Terminology and notation} 
\label{appendix.terminology}

A graph $D = (\mathcal{N}, \mathcal{E})$ consists of a set of nodes, $\mathcal{N}$, and a collection of edges between these nodes, $\mathcal{E}$. The total number of nodes is denoted $p$ and we also refer to this as the \textit{size} of the graph. We generically refer to the nodes as $X_1, ..., X_p$. 

We consider graphs with up to three different types of \textit{edges}. An edge consists of an ordered pair of nodes and an edge type: $(X_i, X_j, e_{ij})$, where $i \neq j$ and $e_{ij} \in \{\leftarrow, \rightarrow, \relbar\}$. We refer to $\relbar$ as \textit{undirected} edges, while the other two types are \textit{directed}. We sometimes furthermore discuss each \textit{endpoint} of the edge type, for example the edge $X_i \to X_j$ is said to have a \textit{tail} at $X_i$ and an \textit{arrowhead} at $X_j$. We will only consider graphs that have at most one edge between any two nodes. 

Two nodes are \textit{adjacent} if there exists an edge between them. The \textit{neighbors} of a node $X_i$, $\text{ne}(X_i)$, are all nodes adjacent to $X_i$. 

A \textit{path} between $X_i$ and $X_r$ is a set of edges 
$$\{(X_i, X_j, e_{ij}), (X_j, X_k, e_{jk}), \dots, (X_q, X_r, e_{qr})\}$$ 
connecting $X_i$ and $X_r$ through consecutive edges. If there exists a path between $X_i$ and $X_r$, we refer to them as being \textit{connected}. 

A \textit{directed path} is a path where all edges have arrowheads, and where all arrowheads point in the same direction. If there exists a directed path from $X_i$ to $X_j$, we refer to $X_i$ as an \textit{ancestor} of $X_j$, and $X_j$ as a \textit{descendant} of $X_i$, denoted $X_i \in \text{an}(X_j)$ and $X_j \in \text{de}(X_i)$, respectively. 

The \textit{parents} of $X_i$ are defined as $\text{pa}(X_i) := \{X_j \mid X_j \in \text{an}(X_i) \text{ and } X_j \in \text{ne}(X_i)\}$, while the \textit{children} of $X_i$ are defined as $\text{ch}(X_i) := \{X_j \mid X_j \in \text{de}(X_i) \text{ and } X_j \in \text{ne}(X_i)\}$.  A \textit{cycle} is a directed path that starts and ends at the same node. 

We consider two types of graphs: 
\begin{enumerate}
\item Directed acyclic graph (DAG): All edges must be directed and no cycles are allowed. 
\item Partially directed acyclic graph (PDAG): No cycles are allowed.
\end{enumerate}
The \textit{skeleton} of a graph can be obtained by removing orientations from all edges. 

In a DAG, two nodes $X_i$ and $X_j$ are said to be $d$-separated by a set of nodes $S \in \mathcal{N} \setminus \{X_i, X_j\}$ if for every path between $X_i$ and $X_j$ at least one of the two following conditions hold: 
\begin{enumerate}
\item $X_k \in S$ and the path includes one of the segments: $X_{m} \leftarrow X_k \rightarrow X_{n}$, $X_{m} \rightarrow X_k \rightarrow X_{n}$, or $X_{m} \leftarrow X_k \leftarrow X_{n}$
\item $X_k, \text{de}(X_k) \notin S$ and $X_{n} \rightarrow X_k \leftarrow X_{k+1}$
\end{enumerate}
We use $X_i \perp_d X_j \mid S$ to denote the corresponding $d$-separation. If $X_{m} \leftarrow X_k \rightarrow X_{n}$, we call $X_k$ a \textit{confounder} on the path. If $X_{m} \rightarrow X_k \leftarrow X_{n}$, we call $X_k$ a \textit{collider} on the path. If the two parents of a collider are not adjacent, we refer to the triple as a \textit{v-structure}. 

We can use a DAG to represent a causal data generating mechanism. Let $\mathbf{X} = X_1, ..., X_p$ be a set of random variables, and for each $X_i$, let $\text{dc}(X_i)$ be direct causes of $X_i$. This corresponds to describing the data generating mechanism through a set of structural equations
$$X_i := f_i(\text{dc}(X_i), \epsilon_i)$$
where $\epsilon_i \Perp (\mathbf{X} \setminus X_i)$ and $f_i$ is some function. If we draw a DAG over $\mathbf{X}$ with an edge from $X_i$ to $X_j$ if and only if $X_i \in \text{dc}(X_j)$, we have that $\text{dc}(X_j) = \text{pa}(X_j)$. We thereby obtain a graphical representation of the causal data generating mechanisms that gives us a useful link between causal relatedness and conditional independence \cite{peters2017}: 
\begin{description}
\item[The Markov property:] If $X_i \perp_d X_j \mid S$ in the DAG, then $X_i \Perp X_j \mid S$ 
\end{description}
We will also assume that the converse of the Markov property is fulfilled:
\begin{description}
\item[Faithfulness assumption:] If $X_i \Perp X_j \mid S$, then $X_i \perp_d X_j$ in the DAG. 
\end{description} 
Note that this assumption is not necessarily fulfilled, even for distributions with densities. However, for such distributions, faithfulness can only be violated on null-sets \cite{pearl2009}. 

\paragraph{CPDAGs} A \textit{Markov equivalence class} contains all DAGs that have the same conditional independence (i.e., Markov) properties. All DAGs in an equivalence class share the same edges and the same v-structures \cite{pearl2009}. Hence, they can be described by a PDAG, where edges are oriented if and only they are oriented in the same direction by all members of the equivalence class, and are otherwise undirected. A PDAG constructed in this way is referred to as a completed PDAG (CPDAG). 

\section{Details about data simulation}
\label{appendix.datasim}

\paragraph{Construction of DAG adjacency matrix}

The DAG adjacency matrices are constructed in the following way:

\begin{enumerate}
\item Let $A$ be a $p \times p$ lower triangular matrix of 1s. Let $n_\text{lowtri} = \frac{p \cdot (p-1)}{2}$ be the number of non-zero entries. 
\item Draw a sparsity $s$ from $\text{Unif}[0,0.8]$.
\item Sample $\lfloor s \cdot n_\text{lowtri}\rceil$ lower triangular entries from $A$ and set these to zero . 
\end{enumerate}

Here, $\lfloor \cdot \rceil$ denotes rounding. The resulting adjacency matrix $A$ will then correspond to a DAG with varying degree of sparsity, as determined by $s$. 

\paragraph{Parameter choices and simulation details}

The standard deviations, $\sigma_i$, are drawn independently from uniform distributions:
$$\sigma_i \sim \text{Unif}[0.5, 2]$$
The regression parameters are constructed as follows (also independently):
$$\beta_{i,j} := b_\text{val} \cdot b_\text{sign}$$
where $b_\text{val} \sim \text{Unif}[0.1, 2]$ and $b_\text{sign}$ is sampled from $\{-1, 1\}$ with probabilities $\{0.4, 0.6\}$, respectively. We use non-uniform sign-sampling to avoid imposing unrealistic symmetry in the simulation setup, which might increase the risk of causal effect estimates almost canceling each other out (empirical non-faithfulness). We do not allow for $\beta_{ij} \simeq 0$ to avoid non-faithfulness. 

The data simulation is performed one variable at a time, noting that the causal ordering of the variables is $(X_1, X_2, ..., X_p)$ due to the lower triangular matrix being used as a starting point for creating the DAG adjacency matrix. 

\paragraph{Constructing features and labels}

Let $\pi$ be a random permutation of the integers $\{1, \dots, p\}$. 

In order to construct the feature data, we compute the Pearson correlation matrix for the full dataset $\mathbf{X} = (X_1, ..., X_p)$. Let $C$ denote this correlation matrix. We then permute both rows and columns of $C$ according to $\pi$ and store the resulting correlation matrix as the feature data. 

In order to obtain the label, we construct the CPDAG corresponding to the Markov equivalence class of the original DAG with adjacency matrix $A$. Let $M$ denote the adjacency matrix corresponding to this CPDAG. We then permute rows and columns of $M$ according to $\pi$ and store the resulting CPDAG adjacency matrix as the label.

\section{Details about the neural network}
\label{appendix.nn}

We use a simple convolutional neural network and  let the number of parameters to depend on $p$. The architecture has the following structure:
\begin{enumerate}
\item Input: $p \times p$ correlation matrix.
\item The input is passed to four parallel convolutional layers, each with $2^5$ filters and with varying kernel sizes and relu activation functions:
\begin{itemize}
\item A $(p, 1)$ kernel (column-wise kernel)
\item A $(1, p)$ kernel (row-wise kernel)
\item A $(1, 1)$ kernel (entry-wise kernel)
\item A $(3, 3)$ kernel (local spatial structure kernel)
\end{itemize}
\item The filters are stacked in a concatenation layer, after which max pooling is applied.
\item The output is flattened and subjected to dropout (with 20\% dropout).
\item A fully connected layer is supplied (relu activation) and 20\% dropout is once again applied.
\item Output: A $p \times p$ matrix of probabilities (using sigmoid activation function).
\end{enumerate}
Full code for specifying the network is available online at \url{https://github.com/annennenne/SLdisco}.

The row-wise and column-wise kernels are intended to make use of the fact that correlation matrices have variables organized in rows and columns. Note that the input is a symmetric matrix. Hence, having both row- and column-wise filters serves in the interest of 1) providing by-variable information in both a row-wise and column-wise manner so it can be combined with other information, and, 2) provide two pathways for utilizing by-variable information hence making this information influence the output more strongly than the entry-wise, and local spatial information. 

This network architecture results in $54,593$ trainable parameters for $p = 5$ node graphs, $1,321,588$ parameters for $p = 10$ nodes and $3,920,528$ parameters for $p = 20$ nodes. 

Note that the specific architecture structure and meta parameters, such as filter sizes and dropout percentages, have been chosen somewhat arbitrarily, and hence this architecture is intended to serve only as a proof of concept example. 

 The models are trained and evaluated on batches of $2^8$ observations. We train the network for 150 epochs (training optimization steps). The weights are optimized using the binary cross-entropy loss function. 

\newpage

\section{Results: Extra figures}
\label{appendix.results.extrafig}

\begin{figure}[h]
\centering
\includegraphics[width = \textwidth]{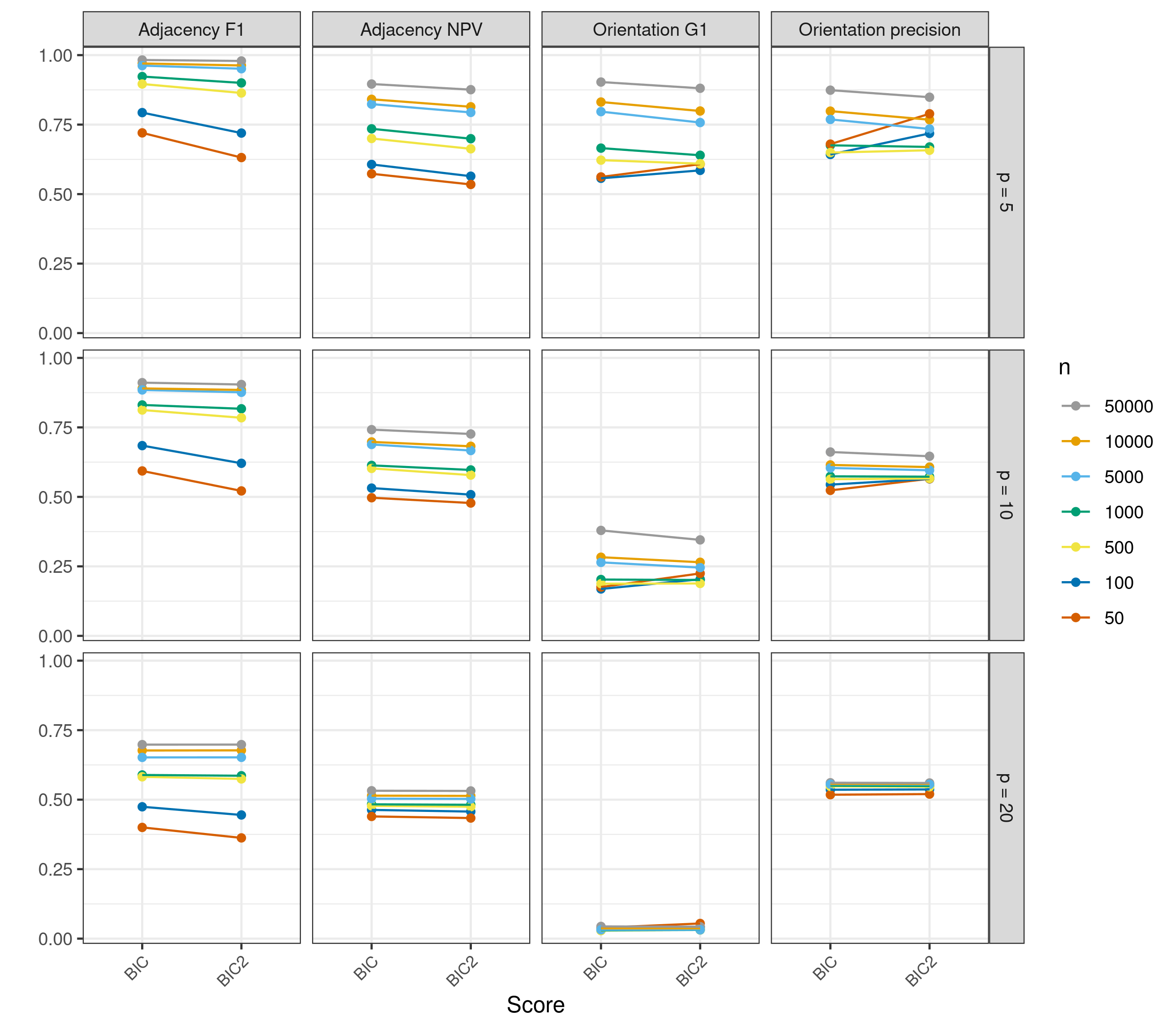}
\caption{Comparison of GES with BIC score and a variation of the BIC score with a larger penalty for the number of estimated parameters. More formally, we write the scores in terms of a penalty $\lambda = \frac{\log(n)}{2}$: Let $\hat{\theta}$ be the Gaussian maximum likelihood estimator and let $l$ denote the negative log-likelihood. Let furthermore $d$ be there number of parameters, i.e. the number of non-zero elements of $\hat\theta$. Then $\text{BIC} = 2 \cdot l(\hat\theta) + \log(n) \cdot d = 2 \cdot  (l(\hat\theta) + \lambda \cdot d)$ and $\text{BIC2} = 2 \cdot l(\hat\theta) + \log(n) \cdot d = 2 \cdot (l(\hat\theta) + 2 \cdot \lambda \cdot d)$. BIC2 thus corresponds to doubling the term that penalizes the number of parameters, which in our use case corresponds to the number of estimated edges.}
\label{fig.ges.bothscores}
\end{figure}

\begin{figure}
\centering
\includegraphics[width = \textwidth]{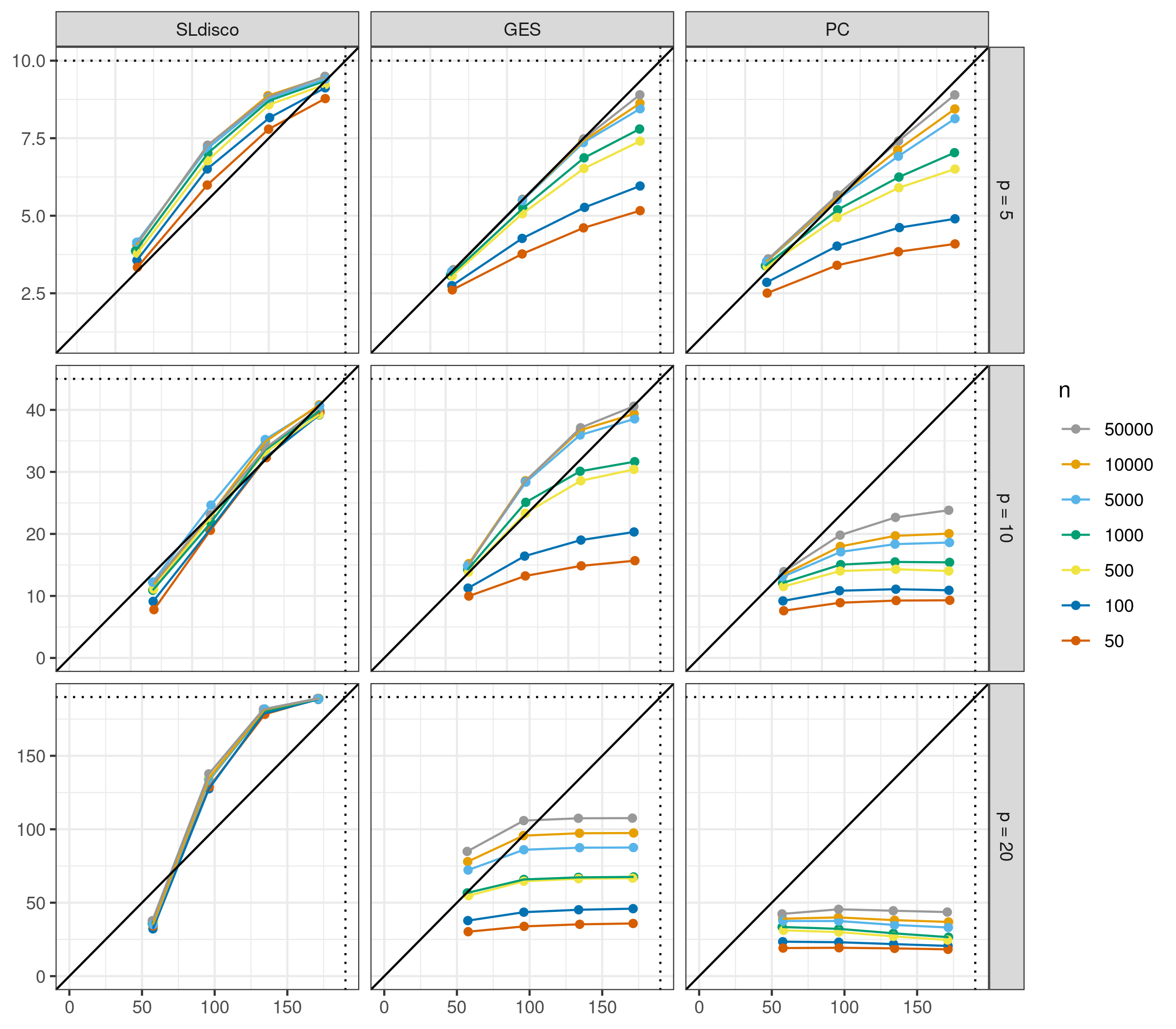}
\caption{Quartiles of estimated and quartiles of true numbers of edges plotted against each other for SLdisco, GES and PC. For SLdisco, the post-processing is conducted using the \textit{BPCO} method with thresholds $\tau = 0.4$ for $p = 5$, $\tau = 0.4$ for $p = 10$ and $\tau = 0.3$ for $p = 20$, following the best approximations found in Section \ref{sec.noedges}. For PC, a significance level of $\alpha = 0.1$ is used. The black line through (0,0) marks perfect edge number estimation, and the dotted lines mark the maximum number of possible edges. The dashed lines marks fully connected graphs, i.e. the maximal number of edges.
We find that PC generally produces graphs that are too sparse. The only case where density is almost correctly estimated is for very large samples and $p = 5$. Generally, the graphs get sparser as the graph size, $p$, increases, and PC fails to recognize denser data generating mechanisms for $p \in \{10, 20\}$, as evident from the almost flat curves. 
GES produces similar results for the $p = 5$ graphs, although with overall a bit better estimation of graph density. For the $p = 10$ graphs, GES estimates graph density somewhat well for $n \geq 500$, but for $p = 20$ graphs, GES struggles to correctly estimate density and, just like PC, does not discriminate well between dense and sparse graphs. 
The results of both PC and GES depends quite strongly on sample size, and especially for small samples, much too sparse models are proposed. 
In contrast, SLdisco is rather agnostic towards the sample size. The average number of edges is approximated well for all graph densities for $p \in \{5, 10\}$, but for $p = 20$ SLdisco produces too sparse solutions for sparser graphs and too dense solutions for denser graphs. It thus seems like the SLdisco for $p = 20$ may be too extreme: Either it produces a very dense graph or a very sparse graph. This may be solvable by better threshold calibration. For example, good approximation of the number of edges for dense graphs (3rd and 4th quartiles) can be achieved by setting $\tau = 0.3$, but in that case, the density of the sparser graphs is more severely underestimated (results not shown).}
\label{fig.nedges.bytruth}
\end{figure}

\newpage 

\section{Application: Extra figures}
\label{appendix.application.extrafig}

\begin{figure}[h]
\centering
\includegraphics[width = \textwidth]{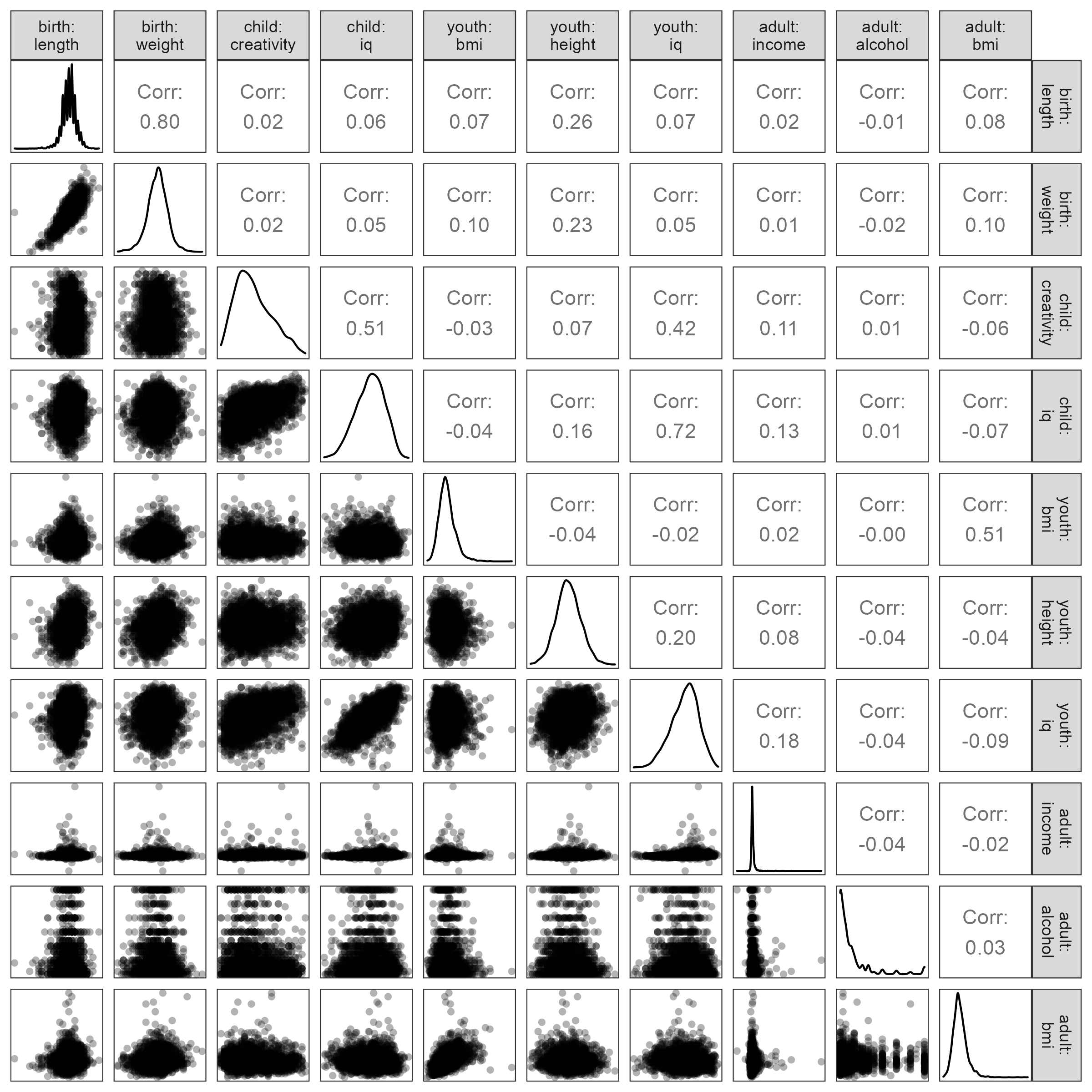}
\caption{Correlations, pairwise and marginal distributions of the 10 Metropolit variables, using the full sample of $n = 2928$ observations.}
\label{fig.appli.plotmatrix}
\end{figure}

\end{document}